\documentclass[12pt]{article}

\usepackage{amssymb}
\usepackage{amstext}

\def\be{\begin{equation}}

\def\ee{\end{equation}}

\def\ba{\begin{eqnarray}}

\def\ea{\end{eqnarray}}

\def \2 {{1 \over 2}}
\def \3 {{1 \over 3}}
\def \4 {{1 \over 4}}
\def \5 {{1 \over 5}}
\def \6 {{1 \over 6}}
\def \7 {{1 \over 7}}
\def \8 {{1 \over 8}}
\def \9 {{1 \over 9}}
\def \0 { \infty}

\begin{document}
\pagestyle{empty}
\begin{flushright}
\begin{tabular}{ll}
MCTP-04-64 & \\
QMUL-PH-04-08 & \\
hep-th/0412027 & \\
02/12/04 & \\ [.1in]
\end{tabular}
\end{flushright}
\begin{center}
{\Large {\bf{Non-associative gauge theory \\
and higher spin interactions}}} \\ [.3in]
{\large{Paul de Medeiros$\, {}^{1}$ and Sanjaye Ramgoolam$\, {}^{2}$}} \\ [.3in]
$\,{}^{1}$\, {\emph{Michigan Center for Theoretical Physics, Randall Laboratory, \\
University of Michigan, Ann Arbor, MI 48109-1120, U.S.A.}} \\ [.2in]
$\,{}^{2}$\, {\emph{Department of Physics, Queen Mary University of London, \\
Mile End Road, London E1 4NS, U.K.}} \\ [.1in]
{\tt{pfdm@umich.edu}} , {\tt{s.ramgoolam@qmul.ac.uk}} \\ [.3in]
{\large{\bf{Abstract}}} \\ [.1in]
\end{center}
We give a framework to describe gauge theory on a certain class of commutative but non-associative fuzzy spaces.
Our description is in terms of an Abelian gauge connection valued in the algebra of functions on the cotangent bundle of the fuzzy space.
The structure of such a gauge theory has many formal similarities with that of Yang-Mills theory.
The components of the gauge connection are functions on the fuzzy space which transform in higher spin representations of the Lorentz group.
In component form, the gauge theory describes an interacting theory of higher spin fields, which remains non-trivial in the limit where the fuzzy space becomes associative.
In this limit, the theory can be viewed as a projection of an ordinary non-commutative Yang-Mills theory.
We describe the embedding of Maxwell theory in this extended framework which follows the standard unfolding procedure for higher spin gauge theories.
\clearpage
\pagestyle{plain}
\pagenumbering{arabic}

%%%%%%%%%%%%%%%%%%%%%%%%%%%%%%%%%%%%%%%%%%%%%%%%%%%%%%%%%%%%%%%%%%%%%%%%%%%%%%%%%%%%%%%%%%%%%%%%%%%%%%%%%%%%%%%%%%%%%%%%%%%%%%%%%%%%%%%%%%%%%%%%%%%%%%%%%%

\section{Introduction}

We formulate gauge theory on a certain class of commutative but non-associative algebras, developing
the constructions initiated in {\cite{ram1}}.
These algebras correspond to so called {\emph{fuzzy spaces}} which reduce to ordinary spacetime manifolds
in a particular associative limit.
We find that such gauge theories have a realisation in terms of interacting higher spin field theories.

The non-associative algebra of interest ${\cal{A}}^*_n (M)$ is a  deformation of the
 algebra of functions
 ${\cal{A}}(M)$ on a $D$-dimensional (pseudo-)Riemannian manifold $M$.
The $*$ denotes a non-associative product for functions on
the fuzzy space whilst $n \in {\mathbb{Z}}_+$ provides a quantitative measure of the
 non-associativity
(in particular ${\cal{A}}^*_\infty = {\cal{A}}$).
For simplicity, we take $M = {\mathbb{R}}^D$ with flat metric.
Most of our formulas will be independent of the signature of this metric, though we
 will take it to be
Lorentzian in discussions of gauge-fixing etc.
Furthermore, although we focus on the deformation for ${\mathbb{R}}^D$, there is a
 conceptually
straightforward generalisation for curved manifolds.
For example, the deformation ${\cal{A}}^*_n ( S^{2k} )$ has been used in the study
 of even-dimensional
 fuzzy spheres in {\cite{ram2}}.

In section 2 we define the commutative, non-associative algebra ${\cal{A}}^*_n ( {\mathbb{R}}^D )$
which deforms ${\cal{A}}( {\mathbb{R}}^D )$, and give the derivations of this algebra.
In this review, we recall that the associator $(A*B)*C - A*(B*C)$ of three functions $A$, $B$ and
 $C$ on ${\cal{A}}^*_n ( {\mathbb{R}}^D )$ can be written as an operator ${\sf{F}} (A,B)$ acting
on $C$ or as an operator ${\sf{E}} (A,C)$ acting on $B$.
These operators have expansions in terms of derivations of the algebra (given in Appendix B) and
 naturally appear when one attempts to construct covariant derivatives for the gauge theory.
We find that an inevitable consequence of this structure is that the connection and gauge parameter
 have to be generalised such that they too have derivative expansions (i.e. they can be understood
as functions on the deformed cotangent bundle ${\cal{A}}^*_n (T^* {\mathbb{R}}^D )$).
The infinite number of component functions in these expansions transform as totally
 symmetric tensors under the Lorentz group.
Consequently we find that this extended gauge theory on
${\cal{A}}^*_n (T^* {\mathbb{R}}^D )$ is related to higher
spin gauge theory on ${\cal{A}}^*_n ( {\mathbb{R}}^D )$.
The local and global structure of this extended gauge theory
is analysed in section 3.

We observe that the extended gauge theory remains non-trivial
even in the limit where the non-associativity parameter goes to zero.
In section 4 we describe certain physical properties in this
associative limit.
In particular we construct a gauge-invariant action and field
equations for the extended theory using techniques related to
the phase space formulation of quantum mechanics initiated by Weyl {\cite{weyl}} and Wigner
 {\cite{wign}}.
The infinite number higher spin components of the extended gauge field become just
tensors on ${\mathbb{R}}^D$ in the associative limit.
We describe various aspects of the extended theory in component
form in order to make the connection with higher spin gauge theory
 more explicit.
From this perspective it will be clear that the extended theory
(as we have presented it) does not realise all the possible symmetries
of the corresponding higher spin theory on ${\mathbb{R}}^D$.
We suggest that it could describe a partially broken phase of some
fully gauge-invariant theory.
We then compare the structure we find with that of the interacting
theory of higher spin fields discovered by Vasiliev {\cite{vas1}}.
A precise way to embed Maxwell theory in the extended theory is given.
The method is identical to the unfolding procedure which has been used
 by Vasiliev in the context of higher spin gauge theories {\cite{vas3}}.
It can also be understood simply via a change of basis in phase space
under a particular symplectic transformation.

In section 5 we describe how the extended theory in the associative limit
described in section 4 is related to a projection of an ordinary
non-commutative Yang-Mills theory. We also describe connections to Matrix theory.
We then discuss how one might generalise the results of section 4 to construct a
gauge-invariant action for the non-associative theory. Section 6 contains some concluding remarks.

%%%%%%%%%%%%%%%%%%%%%%%%%%%%%%%%%%%%%%%%%%%%%%%%%%%%%%%%%%%%%%%%%%%%%%%%%%%%%%%%%%%%%%%%%%%%%%%%%%%%%%%%%%%%%%%%%%%%%%%%%%%%%%%%%%%%%%%%%%%%%%%%%%%%%%%%%%

\section{The non-associative deformation ${\cal{A}}^*_n$}

We begin by defining the non-associative space of interest. Following
{\cite{ram1}}, we consider the commutative, non-associative algebra
${\cal{A}}^*_n ( {\mathbb{R}}^D )$ which is a specific deformation of
the commutative, associative algebra of functions
${\cal{A}} ( {\mathbb{R}}^D )$ on
 ${\mathbb{R}}^D$ (which is to be thought of as physical
spacetime in $D$ dimensions). Another space that will be
important in forthcoming discussions is the algebra of differential operators acting on
${\cal{A}}^*_n ( {\mathbb{R}}^D )$. This algebra is isomorphic to the deformed algebra
${\cal{A}}^*_n ( T^* {\mathbb{R}}^D )$ of functions on
the (flat) cotangent bundle $T^* {\mathbb{R}}^D$. This
correspondence will be helpful when we come to consider
gauge theory on ${\cal{A}}^*_n ( {\mathbb{R}}^D )$.

The space ${\mathbb{R}}^D$ has coordinates $x^\mu$ and flat metric.
The Euclidean signature metric $\delta_{\mu\nu}$ arises most directly in the
Matrix theory considerations motivating \cite{ram1} but the algebra
can be continued to Lorentzian signature by replacing this with Lorentzian metric $\eta_{\mu\nu}$.
The algebraic discussion in this and the next section (and in the appendices) works equally well in either signature, but
some additional subtleties related to gauge-fixing discussed in section 4 are
specific to the Lorentzian case.
The deformed algebra ${\cal{A}}^*_n ( {\mathbb{R}}^D )$ is spanned by
the infinite set of elements $\{ 1, x^\mu , x^{\mu_1 \mu_2} ,... ~  \}$
\footnote{In {\cite{ram1}}, the elements $x$ were called $z$ and
the deformed algebra ${\cal{A}}^*_n ( {\mathbb{R}}^D )$ was called ${\cal{B}}^*_n ( {\mathbb{R}}^D )$.}
, where each $x^{\mu_1 ... \mu_s}$ transforms as a totally symmetric
tensor of rank $s$ under the Lorentz group.
The commutative (but non-associative) product $*$ for all elements $x^{\mu_1 ... \mu_s}$ is
defined in {\cite{ram1}} and Appendix B (this appendix also defines a more general set of products with similar properties to $*$).
The explicit formula is rather
complicated but the important point is that
$x^{\mu_1 ... \mu_s} * x^{\nu_1 ... \nu_t}$ equals
$x^{\mu_1 ... \mu_s \nu_1 ... \nu_t}$ up to the addition of lower rank elements with coefficients proportional to inverse powers
of $n$ (for example $x^\mu * x^\nu = x^\nu * x^\mu = x^{\mu\nu} + {1 \over n} \, \eta^{\mu\nu}$).
This means that the algebra is associative up to terms involving inverse powers of $n$.
An immediate consequence of this structure is that
${\mbox{lim}}_{n \rightarrow \infty} \, {\cal{A}}^*_n ( {\mathbb{R}}^D ) = {\cal{A}} ( {\mathbb{R}}^D )$, since
${\mbox{lim}}_{n \rightarrow \infty} \, x^{\mu_1 ... \mu_s} =
x^{\mu_1} ... x^{\mu_s}$ (the $*$-product being just pointwise
multiplication in this limit). Henceforth we refer to $n \rightarrow \infty$ as
the {\emph{associative limit}}.

Recall that symmetries of non-commutative spaces are typically generated by the subgroup of the
Lorentz group corresponding to symplectic transformations preserving the non-commutativity parameter
(see e.g. {\cite{dounek}}, {\cite{seibwit}}). As explained in {\cite{ram1}} however, ${\cal{A}}^*_n ( {\mathbb{R}}^D )$ corresponds
to a milder deformation of ${\mathbb{R}}^D$ since it is still commutative and the non-associativity parameter
is a Lorentz scalar (proportional to $1/n$). Therefore the deformation above does not break Lorentz symmetry.

One can define derivations $\partial_\mu$ of ${\cal{A}}^*_n
( {\mathbb{R}}^D )$ via the rule
%
%%%     EQUATION (1)
%
\be
\partial_\mu x^{\mu_1 ... \mu_s} \; =\; s\, \delta_\mu^{( \mu_1}
x^{\mu_2 ... \mu_s )} \; ,
\label{eq:1}
\ee
where brackets denote symmetrisation of indices (with weight 1)
\footnote{The derivations $\partial_\mu$ were called $\delta_{\mu}$ in {\cite{ram1}}.}
. This definition implies that $\partial_\mu$ satisfy the Leibnitz rule when
acting on $*$-products of elements of ${\cal{A}}^*_n ( {\mathbb{R}}^D )$.
This Leibnitz property also holds with respect to the more
general commutative, non-associative products described in Appendix B.
 It is clear that composition of these derivations is a commutative
{\emph{and}} associative operation. In the associative $n \rightarrow
\infty$ limit, $\partial_\mu$ just act as the usual partial derivatives
on ${\mathbb{R}}^D$.

%%%%%%%%%%%%%%%%%%%%%%%%%%%%%%%%%%%%%%%%%%%%%%%%%%%%%%%%%%%%%%%%%%%%%%%%%%%%%%%%%%%%%%%%%%%%%%%%%%%%%%%%%%%%%%%%%%%%%%%%%%%%%%%%%%%%%%%%%%%%%%%%%%%%%%%%%%%%

\subsection{Functions}

Functions of the coordinates $x^{\mu_1 ... \mu_s}$ are written $A(x)
 \in {\cal{A}}^*_n ( {\mathbb{R}}^D )$. Such functions form a
commutative but non-associative algebra themselves with respect
to the $*$ multiplication. A quantitative measure of this
non-associativity is given by the {\emph{associator}}
%
%%%     EQUATION (2)
%
\be
[A,B,C] \; :=\; (A * B) * C - A * (B * C)
\label{eq:2}
\ee
for three functions $A$, $B$ and $C$. Since ${\cal{A}}^*_n
( {\mathbb{R}}^D )$ is commutative then the associator
({\ref{eq:2}}) has the antisymmetry $[A,B,C] = - [C,B,A]$.
The associator also satisfies the cyclic identity $[A,B,C] + [B,C,A]
+ [C,A,B] \equiv 0$. An important fact noted in {\cite{ram1}} is that
 such associators can be written as differential operators involving
two functions acting on the third. In particular, one can define the
two operators ${\sf{E}}(A,B)$ and ${\sf{F}}(A,B)$ via
%
%%%     EQUATION (3)
%
\begin{eqnarray}
[A,B,C] &=:&  {\sf{E}}(A,C) \, B \nonumber \\
&=:& {\sf{F}}(A,B) \, C  \; . \label{eq:3}
\end{eqnarray}
The antisymmetry property of the associator implies ${\sf{E}}(A,B)
= -{\sf{E}}(B,A)$ and the cyclic identity implies ${\sf{F}}(A,B) -
{\sf{F}}(B,A) = {\sf{E}}(A,B)$. These operators have the following
derivative expansions (see {\cite{ram1}} or Appendix B)
%
%%%     EQUATION (4)
%
\begin{eqnarray}
{\sf{E}}(A,B) &=& \sum_{s=1}^{\infty} {1 \over s!} \, ( E^{\,\mu_1 ... \mu_s}
 (A,B))(x) \, * \, \partial_{\mu_1} ... \partial_{\mu_s} \; , \nonumber \\
{\sf{F}}(A,B) &=& \sum_{s=1}^{\infty} {1 \over s!} \, ( F^{\,\mu_1 ... \mu_s}
 (A,B))(x) \, * \, \partial_{\mu_1} ... \partial_{\mu_s}  \; , \label{eq:4}
\end{eqnarray}
where the coefficients $E^{\,\mu_1 ... \mu_s} (A,B)$ and
$F^{\,\mu_1 ... \mu_s} (A,B)$ are both polynomial functions of the
 algebra transforming as totally symmetric tensors under the Lorentz group
\footnote{The $*$-product in the expression
 for the operators ({\ref{eq:4}}) means act first on a function with the
derivatives, then $*$-multiply this differentiated function with the
 coefficients (for each $s$ in the sum).}
. The properties quoted above follow for each of these coefficients so
that $E^{\,\mu_1 ... \mu_s} (A,B) = - E^{\,\mu_1 ... \mu_s} (B,A)$ and
$F^{\,\mu_1 ... \mu_s} (A,B) - F^{\,\mu_1 ... \mu_s} (B,A) =
E^{\,\mu_1 ... \mu_s} (A,B)$. The reason there are no $s=0$ terms
 in ({\ref{eq:4}}) is that the associators $[A,1,C]$ and $[A,B,1]$
are both identically zero. Thus since ({\ref{eq:4}}) are valid as
operator equations on any function then including such zeroth order
terms in ({\ref{eq:4}}) would imply their coefficients are identically
zero by simply acting on a constant function. The first non-vanishing
$s=1$ coefficients in ({\ref{eq:4}}) can be expressed rather neatly as
associators, such that $E^{\,\mu} (A,B) = [A, x^\mu ,B]$ and
$F^{\,\mu} (A,B) = [A, B, x^\mu ]$. In a similar manner, all
subsequent $ s > 1 $ coefficients in ({\ref{eq:4}}) can
also be expressed in
terms of (sums of) associators of $A$ and $B$ with coordinates
$x^{\mu_1 ... \mu_s}$ (though we do not give explicit expressions
as they are unnecessary). An important  point to
keep in mind is that  ${\sf{E}}(A,B)$ and ${\sf{F}}(A,B)$ vanish
 in the associative limit as expected.

The algebra of the differential operators in ({\ref{eq:4}}) closes under
composition and is non-associative (following non-associativity
of ${\cal{A}}^*_n ( {\mathbb{R}}^D )$) but it is also non-commutative.
Since ${\sf{E}}(A,B)$ and ${\sf{F}}(A,B)$ vanish in the associative
limit the algebra of these operators becomes trivially
commutative when $n \rightarrow \infty$. As will be seen in the next
 subsection, more general differential operators acting on ${\cal{A}}^*_n ( {\mathbb{R}}^D )$
 also close under composition to form a non-commutative, non-associative
algebra. However, this more general algebra remains non-commutative (but associative) when $n \rightarrow \infty$. For example, the commutator
subalgebra of differential operators acting on ${\mathbb{R}}^D$ corresponding to sections
of the tangent bundle $T {\mathbb{R}}^D$ (i.e. vector fields over
${\mathbb{R}}^D$) is non-Abelian (even though ${\mathbb{R}}^D$ is
 itself commutative). Indeed this is often how one considers simple
non-commutative geometries -- as Hamiltonian phase spaces of ordinary
 commutative position spaces (see e.g. {\cite{dounek}}). We will draw
 on this analogy when we come to construct a gauge theory on
${\cal{A}}^*_n ( {\mathbb{R}}^D )$.

%%%%%%%%%%%%%%%%%%%%%%%%%%%%%%%%%%%%%%%%%%%%%%%%%%%%%%%%%%%%%%%%%%%%%%%%%%%%%%%%%%%%%%%%%%%%%%%%%%%%%%%%%%%%%%%%%%%%%%%%%%%%%%%%%%%%%%%%%%%%%%%%%%%%%%%%%%%%

\subsection{Differential operators}

General differential operators acting on ${\cal{A}}^*_n ( {\mathbb{R}}^D )$ are written
%
%%%     EQUATION (4a)
%
\be
{\hat{A}} \; =\; \sum_{s=0}^{\infty} {1 \over s!} \,
A^{\,\mu_1 ... \mu_s} (x) \, * \, \partial_{\mu_1} ... \partial_{\mu_s} \; ,
\label{eq:4a}
\ee
where $A^{\,\mu_1 ... \mu_s}$ are functions of the algebra
transforming as totally symmetric tensors under the Lorentz group.
It is clear that such operators can equivalently be viewed
as elements of the deformed algebra ${\cal{A}}^*_n ( T^* {\mathbb{R}}^D )$
 of functions on the cotangent bundle $T^* {\mathbb{R}}^D$. In the associative
 limit, ${\cal{A}}^*_\infty ( T^* {\mathbb{R}}^D )$ is just the Weyl algebra
 of ${\mathbb{R}}^D$ (i.e. the infinite-dimensional universal enveloping
algebra of the Heisenberg algebra in $D$-dimensions spanned by all
polynomials of coordinates $x^\mu$ and partial derivatives $\partial_\mu$).
 Because of this  ${\cal{A}}^*_n
(T^* {\mathbb{R}}^D )$  may also be thought of as a {\emph{deformed Weyl algebra}}.

Just as in ({\ref{eq:4}}), the general operators ({\ref{eq:4a}}) also
 close under composition to form a non-commutative and non-associative
 algebra. The operator realisations ${\sf{E}}(A,B)$ and ${\sf{F}}(A,B)$
 ({\ref{eq:3}}) of the associator of functions have useful generalisations
 to the case where functions $A$ and $B$ are replaced by operators
 ${\hat{A}}$ and ${\hat{B}}$ respectively. In particular, we define
%
%%%     EQUATION (4b)
%
\begin{eqnarray}
{\hat{\sf{E}}}({\hat{A}},{\hat{B}}) C &:=& {\hat{B}} ( {\hat{A}} C)
 - {\hat{A}} ( {\hat{B}} C)  \; , \nonumber \\
{\hat{\sf{F}}}({\hat{A}},{\hat{B}}) C &:=& [{\hat{A}},{\hat{B}},C] \; =\; ({\hat{A}}{\hat{B}}) C - {\hat{A}} ( {\hat{B}} C)  \; , \label{eq:4b}
\end{eqnarray}
where $C$ is a function. The definition of ${\hat{\sf{F}}}$ still
involves the associator (just as in ({\ref{eq:3}})). Notice though
 that the definition of ${\hat{\sf{E}}}$ does not involve the associator
 $[{\hat{A}},C,{\hat{B}}] = ({\hat{A}}C){\hat{B}} - {\hat{A}}(C{\hat{B}})$
 directly since the algebra of differential operators is non-commutative. It is the
 ${\hat{\sf{E}}}$ operator defined in ({\ref{eq:4b}}), however, that
will be of interest in the forthcoming discussion. This new definition
 reduces to the usual associator definition ({\ref{eq:3}}) when ${\hat{A}}
 =A$ and ${\hat{B}} =B$ (since the algebra of functions is commutative).
The definitions ({\ref{eq:4b}}) obey the identities ${\hat{\sf{E}}}({\hat{A}}
,{\hat{B}}) \equiv -{\hat{\sf{E}}}({\hat{B}},{\hat{A}})$ and
${\hat{\sf{F}}}({\hat{A}},{\hat{B}}) - {\hat{\sf{F}}}({\hat{B}},
{\hat{A}}) \equiv {\hat{\sf{E}}}({\hat{A}},{\hat{B}}) +
 [{\hat{A}},{\hat{B}}]$ (where $[{\hat{A}},{\hat{B}}] :=
 {\hat{A}}{\hat{B}} - {\hat{B}}{\hat{A}}$ is just the
commutator of operators). These reduce to the identities
found earlier in terms of functions when ${\hat{A}} =A$ and
 ${\hat{B}} =B$. In the associative limit, notice that
${\hat{\sf{F}}}({\hat{A}},{\hat{B}})$ vanishes identically
 whilst ${\hat{\sf{E}}}({\hat{A}},{\hat{B}})$ reduces
 to the commutator $[{\hat{B}},{\hat{A}}]$.

The explicit derivative expansion for
${\hat{\sf{E}}}({\hat{A}},{\hat{B}})$ is given in
 Appendix A for later reference (the corresponding
expression for ${\hat{\sf{F}}}({\hat{A}},{\hat{B}})$
 will not be needed). We should just conclude this review
of the relevant algebras associated with ${\cal{A}}^*_n
 ( {\mathbb{R}}^D )$ by noting that, unlike ({\ref{eq:4}}),
 the operator expression for ${\hat{\sf{E}}}({\hat{A}},{\hat{B}})$
includes a non-vanishing zeroth order algebraic term. It is easy to
 see that this is so by considering $C$ in ({\ref{eq:4b}}) to be the
constant function. In this case all derivative terms in
${\hat{\sf{E}}}({\hat{A}},{\hat{B}})$ on the left hand side vanish
whilst the right hand side reduces to the non-vanishing function
${\hat{B}}A - {\hat{A}}B$ (where $A$ and $B$ are the zeroth order
parts of ${\hat{A}}$ and ${\hat{B}}$ respectively). Thus the zeroth
order part ${\hat{\sf{E}}}({\hat{A}},{\hat{B}})_{(0)} = {\hat{B}}A
- {\hat{A}}B$, which vanishes when ${\hat{A}}=A$ and ${\hat{B}}=B$
as expected.

%%%%%%%%%%%%%%%%%%%%%%%%%%%%%%%%%%%%%%%%%%%%%%%%%%%%%%%%%%%%%%%%%%%%%%%%%%%%%%%%%%%%%%%%%%%%%%%%%%%%%%%%%%%%%%%%%%%%%%%%%%%%%%%%%%%%%%%%%%%%%%%%%%%%%%%%%%%

\section{Non-associative gauge theory}

We begin this section by reviewing the subtleties raised in
{\cite{ram1}} associated with formulating an Abelian gauge
 theory on ${\cal{A}}^*_n ( {\mathbb{R}}^D )$. We show that a
naive formulation is not possible on this non-associative space.
 Instead it is rather natural to consider an extension of such an
Abelian gauge theory on the deformed algebra
${\cal{A}}^*_n ( T^* {\mathbb{R}}^D )$ of functions on  the cotangent bundle.
 We describe the local and global gauge structure of this
non-associative extended theory. We find the structure to be
similar to that of a Yang-Mills theory with infinite-dimensional
 gauge group. We will return to the question of embedding an
 Abelian gauge theory on ${\cal{A}}^*_n ( {\mathbb{R}}^D )$ in
 this extended structure in later sections.

%%%%%%%%%%%%%%%%%%%%%%%%%%%%%%%%%%%%%%%%%%%%%%%%%%%%%%%%%%%%%%%%%%%%%%%%%%%%%%%%%%%%%%%%%%%%%%%%%%%%%%%%%%%%%%%%%%%%%%%%%%%%%%%%%%%%%%%%%%%%%%%%%%%%%%%%%%%

\subsection{Abelian gauge theory on ${\cal{A}}^*_n ( {\mathbb{R}}^D )$}

A necessary ingredient in the construction of any gauge theory is the
 concept of a gauge-covariant derivative. Consider a field $\Phi$
which is a function of ${\cal{A}}^*_n ( {\mathbb{R}}^D )$ and define
it to have the infinitesimal gauge transformation law
%
%%%     EQUATION (5)
%
\be
\delta \Phi \; =\; \epsilon * \Phi \; ,
\label{eq:5}
\ee
where $\epsilon$ is an arbitrary polynomial function of
${\cal{A}}^*_n ( {\mathbb{R}}^D )$. (One reason for the
choice of ({\ref{eq:5}}) is that it is reminiscent of the
infinitesimal gauge transformation for a field in the
 fundamental representation of the gauge group in
ordinary Yang-Mills theory.) An operator $D_\mu$
that is covariant with respect to ({\ref{eq:5}}) must therefore obey
%
%%%     EQUATION (6)
%
\be
\delta ( D_\mu \Phi ) \; =\; \epsilon * ( D_\mu \Phi ) \; .
\label{eq:6}
\ee
Clearly the derivation $\partial_\mu$ ({\ref{eq:1}}) alone does not
obey this covariance requirement since $\delta ( \partial_\mu \Phi ) =
 \epsilon * ( \partial_\mu \Phi ) + ( \partial_\mu \epsilon ) * \Phi$.
To compensate we must introduce a gauge connection $A_\mu$,
 which we take to be a function on ${\cal{A}}^*_n ( {\mathbb{R}}^D )$
 and which transforms such that $\delta ( A_\mu * \Phi ) =
\epsilon * ( A_\mu * \Phi ) - ( \partial_\mu \epsilon ) * \Phi$.
Clearly the existence of such an $A_\mu$ would imply that
%
%%%     EQUATION (7)
%
\be
D_\mu \Phi \; :=\; \partial_\mu \Phi + A_\mu * \Phi
\label{eq:7}
\ee
indeed defines a covariant derivative on functions,
satisfying ({\ref{eq:6}}). Using ({\ref{eq:5}}) then
implies that we require $A_\mu$ to transform such that
%
%%%     EQUATION (8)
%
\be
( \delta A_\mu ) * \Phi \; =\; - ( \partial_\mu \epsilon ) *
\Phi + \epsilon * ( A_\mu * \Phi ) - A_\mu * ( \epsilon * \Phi ) \; .
\label{eq:8}
\ee
In ordinary gauge theory ({\ref{eq:8}}) would allow one to
simply read off the necessary gauge transformation for $A_\mu$
but here things are more complicated due to non-associativity.
 In particular, notice that the last two terms in ({\ref{eq:8}})
 can be written as the associator $[ A_\mu , \Phi , \epsilon ]$
and therefore, using ({\ref{eq:3}}), we require
%
%%%     EQUATION (9)
%
\be
\delta A_\mu  \; =\; - ( \partial_\mu \epsilon ) + {\sf{E}} ( A_\mu , \epsilon ) \; .
\label{eq:9}
\ee
This requirement, however, leads to a contradiction since the
 first two terms in ({\ref{eq:9}}) are algebraic functions on
${\cal{A}}^*_n ( {\mathbb{R}}^D )$ whilst ({\ref{eq:4}}) tells
us that the third term acts only as a differential operator on
${\cal{A}}^*_n ( {\mathbb{R}}^D )$. Therefore such an $A_\mu$
can only exist when ${\sf{E}} ( A_\mu , \epsilon ) =0$, i.e.
 in the associative limit where this would simply be an Abelian
 gauge theory on ${\mathbb{R}}^D$!

As indicated in {\cite{ram1}}, the most conservative way to proceed
 is therefore to simply generalise the gauge connection $A_\mu$ from
 an algebraic function to a differential operator ${\hat{A}}_\mu$ with derivative expansion
%
%%%     EQUATION (10)
%
\be
{\hat{A}}_\mu  \; =\; \sum_{s=0}^{\infty} {1 \over s!} \,
A_{\mu}^{\;\; \alpha_1 ... \alpha_s} (x) \, * \,
\partial_{\alpha_1} ... \partial_{\alpha_s}  \; ,
\label{eq:10}
\ee
where each component $A_{\mu}^{\;\; \alpha_1 ... \alpha_s}$
is an algebraic function on ${\cal{A}}^*_n ( {\mathbb{R}}^D )$
 which transforms in the ($GL(D,{\mathbb{R}})$-reducible) tensor
 product representation corresponding to a vector times a totally
symmetric rank-s tensor of the Lorentz group (we denote this representation
$(1) \otimes (s)$). Unlike the associator operators ({\ref{eq:4}}),
there is no reason not to include all possible terms in
the sum ({\ref{eq:10}}). Indeed, in the associative limit, we will see that the only
algebraic $s=0$ term $A_\mu$ has the interpretation of an
Abelian gauge field embedded in this extended theory. In a
similar manner one can also generalise the gauge parameter
$\epsilon$ to a differential operator ${\hat{\epsilon}}$ with
derivative expansion
%
%%%     EQUATION (11)
%
\be
{\hat{\epsilon}}  \; =\; \sum_{s=0}^{\infty} {1 \over s!} \,
\epsilon^{\alpha_1 ... \alpha_s} (x) \, * \, \partial_{\alpha_1}
 ... \partial_{\alpha_s}  \; .
\label{eq:11}
\ee

As noted already, the algebra of such operators is both
non-associative and non-commutative. Consequently we must take
 care when revising the arguments of this subsection in terms
of these extended fields. This revised analysis is
described, in the next subsection,
within the framework of global gauge transformations for
the extended theory.

In concluding, it is important to stress that the generalisation
we have made is a modification of the original theory and therefore
the extended theory need not trivially reduce to an Abelian gauge
theory on ${\mathbb{R}}^D$ in the associative limit. (Notice that
 the $s>0$ terms in ({\ref{eq:10}}) and ({\ref{eq:11}}) do not vanish
as $n \rightarrow \infty$.)  Indeed we will find it does not though
we will give a precise way to embed the Abelian theory in its extension
on ${\mathbb{R}}^D$.

%%%%%%%%%%%%%%%%%%%%%%%%%%%%%%%%%%%%%%%%%%%%%%%%%%%%%%%%%%%%%%%%%%%%%%%%%%%%%%%%%%%%%%%%%%%%%%%%%%%%%%%%%%%%%%%%%%%%%%%%%%%%%%%%%%%%%%%%%%%%%%%%%%%%%%%%%%%%

\subsection{Global structure}

Consider again a field $\Phi$ which is a function of
${\cal{A}}^*_n ( {\mathbb{R}}^D )$ but now with
infinitesimal gauge transformation law
%
%%%     EQUATION (12)
%
\be
\delta \Phi \; =\; {\hat{\epsilon}}\, \Phi \; ,
\label{eq:12}
\ee
where ${\hat{\epsilon}}$ is the extended differential operator
({\ref{eq:11}}). Formally this is similar to Yang-Mills theory
 where one then obtains the global gauge transformation by exponentiating
the local (Lie algebra valued) gauge parameter to obtain a general
Lie group element (or more precisely the fundamental representations
of these quantities). The main difference here is that the algebra of
local gauge transformations ({\ref{eq:12}}) is non-associative. Despite
this, given a general differential operator ${\hat{\epsilon}}$, there still
 exists a well-defined exponential ${\mbox{exp}} ( {\hat{\epsilon}} )$
 {\cite{nessab}}. The construction essentially just follows the power
series definition of the exponential map for matrix algebras but here
 one must choose an ordering for powers of ${\hat{\epsilon}}$
(so as to avoid the potential ambiguities due to non-associativity).
We follow {\cite{nessab}} and define powers via a \lq left action'
rule so that
%
%%%     EQUATION (12a)
%
\be
{\mbox{exp}} ( {\hat{\epsilon}} ) \, \Phi \; :=\; \Phi + {\hat{\epsilon}}\,
 \Phi + {1 \over 2} \, {\hat{\epsilon}} \, ( {\hat{\epsilon}}\, \Phi ) +
{1 \over 3!} \, {\hat{\epsilon}} \, ( {\hat{\epsilon}} \,
( {\hat{\epsilon}}\, \Phi ))+ .\, .\, . \;\; ,
\label{eq:12a}
\ee
for any function $\Phi$. It is then clear that the exponentiated
operator ${\hat{g}} := {\mbox{exp}} ( {\hat{\epsilon}} )$ is also
 a differential operator acting on the algebra (albeit a rather complicated
function of ${\hat{\epsilon}}$) and we define the \lq global'
transformation of $\Phi$ to be
%
%%%     EQUATION (13)
%
\be
\Phi \; \rightarrow \; {\hat{g}} \, \Phi \; .
\label{eq:13}
\ee
This transformation obviously reduces to ({\ref{eq:12}}) in
some neighbourhood of the identity where ${\hat{g}} = 1 +
{\hat{\epsilon}}$ (the \lq identity' here is the unit element
 of ${\cal{A}}^*_n ( {\mathbb{R}}^D )$). The set of all transformations
({\ref{eq:13}}) does not quite form a group under left action composition
 since it fails to satisfy the associativity axiom (due to
 non-associativity of the algebra). However, all the other group axioms
 are satisfied
\footnote{Closure under composition follows using an extension of
the Baker-Campbell-Hausdorff formula whereby the composition of two
 exponentials can itself be expressed as an exponential with the
 exponent being the sum of the two original exponents plus corrections
involving commutators and associators of these exponents. As already
mentioned above, the identity element is simply the unit element of
${\cal{A}}^*_n ( {\mathbb{R}}^D )$. Every element ${\hat{g}} =
{\mbox{exp}} ( {\hat{\epsilon}} )$ has the left inverse ${\hat{g}}^{-1}
:= {\mbox{exp}} (- {\hat{\epsilon}} )$ which satisfies ${\hat{g}}^{-1}
 ( {\hat{g}} \Phi ) = {\hat{g}} ( {\hat{g}}^{-1} \Phi ) = \Phi$ for any
 function $\Phi$. Thus ${\hat{g}}^{-1} = 1 - {\hat{\epsilon}}$ locally.}
.

The derivation $\partial_\mu$ is not covariant with respect ({\ref{eq:13}})
since this transformation implies $\partial_\mu \Phi \rightarrow [ \partial_\mu ,
{\hat{g}} ] \Phi + {\hat{g}} ( \partial_\mu \Phi )$. As noted at the end of
the previous subsection, we therefore introduce a gauge connection
${\hat{A}}_\mu$
which must transform such that ${\hat{A}}_\mu \Phi \rightarrow -[ \partial_\mu ,
 {\hat{g}} ] \Phi + {\hat{g}} ( {\hat{A}}_\mu \Phi )$ in order that
%
%%%     EQUATION (14)
%
\be
{\hat{D}}_\mu \Phi \; :=\; \partial_\mu \Phi + {\hat{A}}_\mu \Phi
\label{eq:14}
\ee
transforms covariantly under ({\ref{eq:13}}). This necessary gauge
 transformation of ${\hat{A}}_\mu \Phi$ under ({\ref{eq:13}}) can be
realised provided the gauge transformation of ${\hat{A}}_\mu$ is
defined such that
%
%%%     EQUATION (15)
%
\be
{\hat{A}}_\mu \Phi \; \rightarrow \; - [ \partial_\mu , {\hat{g}} ]
( {\hat{g}}^{-1} \Phi^\prime ) + {\hat{g}} ( {\hat{A}}_\mu
( {\hat{g}}^{-1} \Phi^\prime ))
\label{eq:15}
\ee
under the more general function transformation $\Phi \rightarrow
 \Phi^\prime$. This gives the desired gauge transformation when
 $\Phi^\prime = {\hat{g}} \, \Phi$. One can obtain the gauge
transformation of ${\hat{A}}_\mu$ itself by using the operator
${\hat{\sf{F}}}$ ({\ref{eq:4b}}) to rearrange the brackets in
({\ref{eq:15}}). In particular, notice that the right hand side
of ({\ref{eq:15}}) can be written
%
%%%     EQUATION (16)
%
\begin{eqnarray}
&& \left( - [ \partial_\mu , {\hat{g}} ] + {\hat{g}} {\hat{A}}_\mu -
{\hat{\sf{F}}} ({\hat{g}},{\hat{A}}_\mu ) \right) ( {\hat{g}}^{-1}
 \Phi^\prime ) \label{eq:16} \\
&&=\; \left( \left( - [ \partial_\mu , {\hat{g}} ] + {\hat{g}} {\hat{A}}_\mu
- {\hat{\sf{F}}} ({\hat{g}},{\hat{A}}_\mu ) \right) {\hat{g}}^{-1}
\right)
\Phi^\prime - {\hat{\sf{F}}} \left( \left( - [ \partial_\mu , {\hat{g}} ] +
{\hat{g}} {\hat{A}}_\mu - {\hat{\sf{F}}} ({\hat{g}},{\hat{A}}_\mu ) \right)
 , {\hat{g}}^{-1} \right) \Phi^\prime   \; . \nonumber
\end{eqnarray}
Therefore ${\hat{A}}_\mu$ must have the following gauge transformation
%
%%%     EQUATION (17)
%
\be
{\hat{A}}_\mu \; \rightarrow \; \left( - [ \partial_\mu , {\hat{g}} ] + {
\hat{g}} {\hat{A}}_\mu - {\hat{\sf{F}}} ({\hat{g}},{\hat{A}}_\mu )
\right) {\hat{g}}^{-1} - {\hat{\sf{F}}} \left( \left( - [ \partial_\mu ,
{\hat{g}} ] + {\hat{g}} {\hat{A}}_\mu - {\hat{\sf{F}}} ({\hat{g}},
{\hat{A}}_\mu ) \right) , {\hat{g}}^{-1} \right) \; .
\label{eq:17}
\ee
Setting ${\hat{g}} = 1 + {\hat{\epsilon}}$ in ({\ref{eq:17}})
leads to the infinitesimal form of the gauge transformation
%
%%%     EQUATION (18)
%
\be
\delta {\hat{A}}_\mu \; = \;  - [ \partial_\mu , {\hat{\epsilon}} ] +
{\hat{\sf{E}}} ( {\hat{A}}_\mu ,{\hat{\epsilon}} )  \; .
\label{eq:18}
\ee
Of course, at the infinitesimal level, this transformation equivalently follows by the
requirement that $\delta ( {\hat{D}}_{\mu} \Phi ) = {\hat{\epsilon}} ( {\hat{D}}_{\mu} \Phi )$
under ({\ref{eq:12}}).

Notice that ({\ref{eq:17}}) and ({\ref{eq:18}}) do not quite take
the form one would expect by naively following the Yang-Mills analogy
(that is they differ from what one might expect by associator terms).
This is  a consequence of the non-associativity of the  underlying algebra of functions.
 In the following section we will find that the
expected Yang-Mills type structure follows exactly in the associative limit.

In the discussion above we
 have only defined covariant derivatives ${\hat{D}}_\mu$ on functions
 and not on differential operators. Although not of the standard Yang-Mills form,
 (minus) the right hand side of ({\ref{eq:18}}) can still be taken as
the definition for the action of the covariant derivative on operator ${\hat{\epsilon}}$, such that
%
%%%     EQUATION (18a)
%
\be
{\hat{D}}_\mu \cdot {\hat{\epsilon}} \; := \; [ \partial_\mu ,
{\hat{\epsilon}} ] + {\hat{\sf{E}}} ( {\hat{\epsilon}}, {\hat{A}}_\mu )   \; .
\label{eq:18a}
\ee
This statement is partially justified by the fact that
${\hat{D}}_\mu$ then satisfies the Leibnitz rule ${\hat{D}}_\mu
( {\hat{\epsilon}} \Phi ) = ( {\hat{D}}_\mu \cdot {\hat{\epsilon}} )
 \Phi + {\hat{\epsilon}} ( {\hat{D}}_\mu \Phi )$ (for general
operator ${\hat{\epsilon}}$ and function $\Phi$)
\footnote{It should be noted that the  naive commutator action $[ \partial_\mu +
{\hat{A}}_\mu , {\hat{\epsilon}} ]$
 on an operator
${\hat{\epsilon}}$ fails to satisfy the Leibnitz
rule due to associator terms. The commutator action
is identical to the covariant derivative proposed above in the
associative limit.}
.

Based on the transformation law found above, we define the
{\emph{field strength}} ${\hat{F}}_{\mu\nu}$ as
%
%%%     EQUATION (19)
%
\be
{\hat{F}}_{\mu\nu} \; := \;  {\hat{\sf{E}}} ( {\hat{D}}_\nu ,
{\hat{D}}_\mu ) \; =\; [ \partial_\mu , {\hat{A}}_\nu ] - [ \partial_\nu ,
 {\hat{A}}_\mu ]  + {\hat{\sf{E}}} ( {\hat{A}}_\nu , {\hat{A}}_\mu )  \; .
\label{eq:19}
\ee
It is clear from this definition that ${\hat{F}}_{\mu\nu}$ is indeed
a differential operator which transforms as a two-form under
 the Lorentz group. In addition, since the gauge transformations above
imply that
%
%%%     EQUATION (20)
%
\be
{\hat{D}}_\mu \Phi \; \rightarrow \; {\hat{g}} \, ( {\hat{D}}_\mu
( \, {\hat{g}}^{-1} \, \Phi^\prime ))  \; ,
\label{eq:20}
\ee
under ({\ref{eq:15}}), then it follows that ${\hat{F}}_{\mu\nu} \Phi =
 {\hat{D}}_\mu ( {\hat{D}}_\nu \Phi ) - {\hat{D}}_\nu ( {\hat{D}}_\mu
\Phi )$ transforms as
%
%%%     EQUATION (21)
%
\be
{\hat{F}}_{\mu\nu} \Phi \; \rightarrow \; {\hat{g}} \, (
 {\hat{F}}_{\mu\nu} ( \, {\hat{g}}^{-1} \, \Phi^\prime ))  \; ,
\label{eq:21}
\ee
and is therefore also gauge-covariant when $
\Phi^\prime = {\hat{g}} \, \Phi$. The infinitesimal form of
the covariant gauge transformation of ${\hat{F}}_{\mu\nu}$ is
%
%%%     EQUATION (22)
%
\be
\delta {\hat{F}}_{\mu\nu} \; =\; {\hat{\sf{E}}} ( {\hat{F}}_{\mu\nu}
, {\hat{\epsilon}} ) \; .
\label{eq:22}
\ee

From the evidence above, it is clear that there are various subtleties
related to the non-associative nature of the theory.
Indeed the non-associativity complicates matters even further in the
description of more physical aspects of the theory like Lagrangians,
field equations and the embedding of an Abelian gauge theory in this
 extended framework.
Recall though that this extended theory should have a non-trivial
structure, even in the associative limit.
We therefore postpone further discussion of the non-associative
extended theory to analyse its associative limit in more detail.

%%%%%%%%%%%%%%%%%%%%%%%%%%%%%%%%%%%%%%%%%%%%%%%%%%%%%%%%%%%%%%%%%%%%%%%%%%%%%%%%%%%%%%%%%%%%%%%%%%%%%%%%%%%%%%%%%%%%%%%%%%%%%%%%%%%%%%%%%%%%%%%%%%%%%%%%%%%

\section{Gauge theory on $T^* {\mathbb{R}}^D$ and higher spin gauge theory on ${\mathbb{R}}^D$}

We begin this section by briefly summarising the results of the
previous subsection in the associative limit.
We then describe how one can construct a gauge-invariant action
and equations of motion for this theory.
Writing the extended gauge field ${\hat{A}}_\mu$ in terms of
component functions $A_{\mu}^{\;\; \alpha_1 ... \alpha_s}$ we find
that the extended theory describes an interacting
 theory involving an infinite number of higher spin fields.
When written in component form, it will be clear that the extended
theory (as we have described it) does not realise all the possible
symmetries of the corresponding higher spin gauge theory.
We suggest that the extended theory could correspond to a
 partially broken phase of some fully gauge-invariant higher
spin theory.
A comparison of the structure we find with that of the
interacting theory of higher spin fields discovered by
Vasiliev {\cite{vas1}} is then given.
We conclude the section by showing how an Abelian gauge
theory can be embedded in this extended framework.
The embedding is related to the unfolding procedure used
by Vasiliev in the context of higher spin gauge theory {\cite{vas3}}.

%%%%%%%%%%%%%%%%%%%%%%%%%%%%%%%%%%%%%%%%%%%%%%%%%%%%%%%%%%%%%%%%%%%%%%%%%%%%%%%%%%%%%%%%%%%%%%%%%%%%%%%%%%%%%%%%%%%%%%%%%%%%%%%%%%%%%%%%%%%%%%%%%%%%%%%%%%%

\subsection{The associative limit}

Many expressions found in the previous section retain their
schematic form in the associative limit. For example, the
gauge transformations for functions are just as in ({\ref{eq:12}}),
 ({\ref{eq:13}}) though $\Phi$ is now simply a function on ${\mathbb{R}}^D$
 whilst operators like ${\hat{\epsilon}}$ in ({\ref{eq:11}}) now have
the expansion
%
%%%     EQUATION (22a)
%
\be
{\hat{\epsilon}}  \; =\; \sum_{s=0}^{\infty} {1 \over s!} \,
\epsilon^{\alpha_1 ... \alpha_s} (x) \, \partial_{\alpha_1} ...
\partial_{\alpha_s}  \; ,
\label{eq:22a}
\ee
in terms of an infinite number of functions
$\epsilon^{\alpha_1 ... \alpha_s}$ on ${\mathbb{R}}^D$
(which still transform as totally symmetric tensors under the Lorentz group). It should
also be noted that the set of gauge transformations ({\ref{eq:13}})
 now form a group since the associativity axiom is no longer violated
in this limit.

Recall that the associator operators ${\sf{E}}$, ${\sf{F}}$ and
${\hat{\sf{F}}}$ in ({\ref{eq:4}}), ({\ref{eq:4b}}) vanish in
 the associative limit whilst ${\hat{\sf{E}}} ({\hat{A}},{\hat{B}})$
 reduces to the commutator $[ {\hat{B}},{\hat{A}} ]$. Consequently
the gauge transformation ({\ref{eq:17}}) takes the more familiar
 Yang-Mills form
%
%%%     EQUATION (23)
%
\be
{\hat{A}}_\mu \; \rightarrow \; - [ \partial_\mu , {\hat{g}} ] \,
{\hat{g}}^{-1} + {\hat{g}} \, {\hat{A}}_\mu \, {\hat{g}}^{-1}  \; .
\label{eq:23}
\ee
This reduces to the infinitesimal variation
%
%%%     EQUATION (24)
%
\be
\delta {\hat{A}}_\mu \; = \;  - [ \partial_\mu + {\hat{A}}_\mu ,
{\hat{\epsilon}} ]  \; .
\label{eq:24}
\ee
As explained earlier, from this transformation we define the action
 of the covariant derivative ${\hat{D}}_\mu = \partial_\mu +
{\hat{A}}_\mu$ on operators ${\hat{\epsilon}}$ to be ${\hat{D}}_\mu
\cdot {\hat{\epsilon}} := [ {\hat{D}}_\mu , {\hat{\epsilon}} ]$.

Since connections and gauge parameters are functions of both $x$ and $\partial$
then the associative limit of the extended theory in $D$ dimensions is also related to Yang-Mills
theory on a $2D$-dimensional non-commutative space. This connection will be clarified in section 5.1.

The transformation ({\ref{eq:23}}) implies that ${\hat{D}}_\mu$
indeed transforms covariantly as
%
%%%     EQUATION (25)
%
\be
{\hat{D}}_\mu \; \rightarrow \; {\hat{g}} \, {\hat{D}}_\mu \,
{\hat{g}}^{-1}  \; .
\label{eq:25}
\ee
Hence the field strength ({\ref{eq:19}})
%
%%%     EQUATION (26)
%
\be
{\hat{F}}_{\mu\nu} \; = \; [ {\hat{D}}_\mu , {\hat{D}}_\nu ] \; =\;
[ \partial_\mu , {\hat{A}}_\nu ] - [ \partial_\nu , {\hat{A}}_\mu ]
+ [ {\hat{A}}_\mu , {\hat{A}}_\nu ]  \; ,
\label{eq:26}
\ee
also transforms covariantly. The infinitesimal form of this covariant
 transformation being
%
%%%     EQUATION (27)
%
\be
\delta {\hat{F}}_{\mu\nu} \; =\; [ {\hat{\epsilon}} , {\hat{F}}_{\mu\nu} ] \; .
\label{eq:27}
\ee
%

%%%%%%%%%%%%%%%%%%%%%%%%%%%%%%%%%%%%%%%%%%%%%%%%%%%%%%%%%%%%%%%%%%%%%%%%%%%%%%%%%%%%%%%%%%%%%%%%%%%%%%%%%%%%%%%%%%%%%%%%%%%%%%%%%%%%%%%%%%%%%%%%%%%%%%%%%%%

\subsection{Action and field equations}

A simple equation of motion to consider for the extended theory in the associative limit is
%
%%%     EQUATION (28)
%
\be
[ {\hat{D}}^\mu , {\hat{F}}_{\mu\nu} ] \; =\; 0 \; .
\label{eq:28}
\ee
This is  the field equation one would expect from
following the Yang-Mills type structure found for the extended
theory in the previous subsection. The equation ({\ref{eq:28}})
 is invariant under the gauge transformation ({\ref{eq:23}}).
Moreover it is this equation (rather than, say, the also
gauge-invariant equation ${\hat{D}}^\mu {\hat{F}}_{\mu\nu} =0$)
which reduces to the correct Maxwell equation as we will see in section 4.5.

Following the Yang-Mills analogy further, a natural gauge-invariant action to consider is of the form
%
%%%     EQUATION (29)
%
\be
-{1 \over 4}  {\mbox{Tr}} \left( {\hat{F}}_{\mu\nu} {\hat{F}}^{\mu\nu}
 \right) \; .
\label{eq:29}
\ee
Such an action can be constructed for the extended theory we are
considering provided there exists a well-defined map
%
%%%     EQUATION (30)
%
\be
 {\mbox{Tr}} \; : \; {\cal{A}} ( T^* {\mathbb{R}}^D ) \; \rightarrow
\; {\mathbb{R}} \; .
\label{eq:30}
\ee
Furthermore, since we are now dealing with an associative theory,
it is clear that an action ({\ref{eq:29}}) would be gauge-invariant
 provided the map ({\ref{eq:30}}) is symmetric, such that it satisfies
%
%%%     EQUATION (31)
%
\be
 {\mbox{Tr}} ( {\hat{A}} {\hat{B}} ) \; = \;  {\mbox{Tr}}
( {\hat{B}} {\hat{A}} ) \; ,
\label{eq:31}
\ee
for any differential operators ${\hat{A}}$ and ${\hat{B}}$.
The next task is therefore to show that such a symmetric map exists.

Before going into the details of the map we should make a few remarks.
Firstly, notice that we write the map ${\mbox{Tr}}$ which alludes to the
Yang-Mills analogy where it simply consists of taking the usual
gauge-invariant trace (using the Cartan-Killing metric for the gauge
 group) followed by integrating over spacetime. However, we do not
assume a priori that the map ({\ref{eq:30}}) can be factorised in
 this way
\footnote{As explained in {\cite{dounek}}, non-commutative gauge theories
 provide a counter example where such a factorisation of ${\mbox{Tr}}$
is not possible.}
. In the Yang-Mills case the symmetry property of ${\mbox{Tr}}$ simply
 follows from the fact that the trace is symmetric. The symmetry of the
 trace is a rather general property of finite-dimensional representations
-- as one considers for Yang-Mills theories with compact gauge groups --
since such representations can be expressed in terms of finite-dimensional
square matrices (and for two such matrices $X$, $Y$, the trace of $XY$ is
 just $X^{i}_{\;\; j} Y^{j}_{\;\; i} = Y^{i}_{\;\; j} X^{j}_{\;\; i}$).
For the extended theory we are considering though fields are valued in the
algebra of differential operators on ${\mathbb{R}}^D$ and the
 situation is very different for the case of such infinite-dimensional
representations. For example, in quantum mechanics, if the Heisenberg
 algebra $[{\hat{x}},{\hat{p}}] =i$ had any representations of finite
 dimension $n \neq 0$ (and hence a symmetric trace) then it would
imply the well-known contradiction $0=i n$ !

The example above is quite pertinent since we will now show that
 fields in the extended theory we are considering are related to certain
functions in the formulation of quantum mechanics based on
the original work of Weyl {\cite{weyl}} and Wigner {\cite{wign}} which was
 later developed by Groenewold {\cite{groe}} and Moyal {\cite{moya}}
(see {\cite{zach}} for a nice review). Within this framework,
there exists a natural concept of the symmetric map ${\mbox{Tr}}$.
In terms of the abstract canonically conjugate operators ${\hat{x}}^\mu$
 and ${\hat{p}}_\mu$, a general operator ${\hat{A}}$ of the form
({\ref{eq:22a}}) is written
%
%%%     EQUATION (32)
%
\be
{\hat{A}} \; =\; A ({\hat{x}},{\hat{p}})  \; =\; \sum_{s=0}^{\infty}
 {i^s \over s!} \, A^{\alpha_1 ... \alpha_s} ({\hat{x}}) \,
{\hat{p}}_{\alpha_1} ... {\hat{p}}_{\alpha_s} \; .
\label{eq:32}
\ee
Such objects form the most general set of operators for a
quantum mechanical system on ${\mathbb{R}}^D$. These operators
can be given definite Hermiticity properties by simply reordering
 ${\hat{x}}$ and ${\hat{p}}$ appropriately in ({\ref{eq:32}}) at
the expense of changing the values of the coefficients in the
expansion of a general $A ({\hat{x}},{\hat{p}})$. Of course this
would put restrictions on the kind of coefficient functions
$A^{\alpha_1 ... \alpha_s}$ permitted in ({\ref{eq:32}}). Let
us therefore just proceed with the \lq ${\hat{p}}$ to the right'
 ordering prescription above
\footnote{This prescription is sometimes referred to as {\emph{standard ordering}} and was considered originally by Mehta {\cite{meht}}.
An alternative prescription (considered by {\cite{zach}}
and the references therein) is the totally symmetric {\emph{Weyl ordering}}
 of ${\hat{x}}$ and ${\hat{p}}$ in a given operator. Weyl
ordering guarantees that operators are Hermitean.
We are very grateful to C. Zachos for pointing out the comprehensive review by Lee {\cite{lee}}
of the precise relationships between the various possible ordering prescriptions.}
.

Given this ordering rule, the {\emph{Weyl homomorphism}} {\cite{weyl}}
says that every operator $A({\hat{x}},{\hat{p}})$ ({\ref{eq:32}}) is
 naturally associated with an ordinary c-number function ${\tilde{A}}$
 on the classical phase space ${\mathbb{R}}^{2D}$ (spanned by coordinates
$( x , p )$), such that
%
%%%     EQUATION (33)
%
\be
A({\hat{x}},{\hat{p}})  \; =\; {1 \over (2 \pi )^{2D}} \int dy\,  dq\, dx\,
dp \; {\tilde{A}} (x,p) \; {\mbox{exp}} (i\, q_\mu ( {\hat{x}}^\mu - x^\mu ))
 \; {\mbox{exp}} ( i\, y^\mu ( {\hat{p}}_\mu - p_\mu )) \; .
\label{eq:33}
\ee
The operator ${\hat{A}}$ and function ${\tilde{A}}$ in ({\ref{eq:33}}) are
then said to be {\emph{Weyl-dual}}. For the ordering rule we have chosen,
notice that the $q$ and $y$ integrals can be evaluated in ({\ref{eq:33}})
to give the formal Dirac delta functions
$\delta ( {\hat{x}} - x )$ and $\delta ( {\hat{p}} - p )$ respectively.
Since these delta functions involve the operators ${\hat{x}}$ and
${\hat{p}}$,   they do not commute. Their arrangement in ({\ref{eq:33}})
 clearly respects our ordering rule. Therefore, for example, if ${\tilde{A}}$
 is a polynomial function of the phase space variables $x$ and $p$ then
({\ref{eq:33}}) says that the corresponding operator function $A$ is exactly
the same polynomial function, but of the operators ${\hat{x}}$ and
${\hat{p}}$ respectively -- with the ${\hat{p}}$ operators ordered
to the right. In particular this means that ({\ref{eq:33}}) relates
the operators ${\hat{x}}^\mu$ and ${\hat{p}}_\nu$ to the classical
phase space coordinates  $x^\mu$ and $p_\nu$ respectively. The
coefficient position operators $A^{\alpha_1 ... \alpha_s} ({\hat{x}})$
in ({\ref{eq:32}}) can therefore also be written in terms of the
function ${\tilde{A}}$, such that
%
%%%     EQUATION (33a)
%
\be
A^{\alpha_1 ... \alpha_s} ({\hat{x}})  \; =\; {1 \over (2 \pi )^{2D}}
\int dy\,  dq\, dx\, dp \; {\tilde{A}} (x,p) \; y^{\alpha_1} .\, .\, .\,
 y^{\alpha_s} \; {\mbox{exp}} (i\, q_\mu ( {\hat{x}}^\mu - x^\mu ) -i\,
y^\mu p_\mu )  \; .
\label{eq:33a}
\ee

The trace ${\mbox{Tr}}$ of the operator $A({\hat{x}},{\hat{p}})$ is
 defined by
%
%%%     EQUATION (34)
%
\be
 {\mbox{Tr}} \, ( {\hat{A}} )  \; :=\; \int dx\, dp \; {\tilde{A}} (x,p) \; .
\label{eq:34}
\ee
This integral is only defined for functions ${\tilde{A}}$
with suitably rapid asymptotic decay properties. We will describe a
 particular {\emph{Wigner basis}} for a class of  such integrable functions in
the next subsection.

The inverse of the relation ({\ref{eq:33}}) can then be expressed in
terms of this trace, such that
%
%%%     EQUATION (35)
%
\be
{\tilde{A}} (x,p) \; =\; {1 \over (2 \pi )^{2D}} \int dy\,  dq \;\,
{\mbox{exp}} ( i\, ( q_\mu x^\mu + y^\mu p_\mu ) )  \; {\mbox{Tr}}
\left( {\mbox{exp}} ( -i\, q_\mu {\hat{x}}^\mu ) \; {\hat{A}} \;
{\mbox{exp}} ( -i\, y^\mu {\hat{p}}_\mu ) \right) \; .
\label{eq:35}
\ee

That ({\ref{eq:33}}) defines a homomorphism was first noted by von
 Neumann {\cite{vneu}}. It follows from the fact that, given two
operators ${\hat{A}}$ and ${\hat{B}}$ (of the form ({\ref{eq:32}}))
with respective Weyl-dual functions ${\tilde{A}}$ and ${\tilde{B}}$,
then one can find a new function denoted ${\tilde{A}}\star {\tilde{B}}$
which is related to the operator product ${\hat{A}} {\hat{B}}$ precisely
as in ({\ref{eq:33}}) (i.e. ${\hat{A}} {\hat{B}}$ and ${\tilde{A}}
\star {\tilde{B}}$ are also Weyl-dual)
\footnote{In general, the product ${\hat{A}} {\hat{B}}$ of two
 ordered operators ${\hat{A}}$ and ${\hat{B}}$ is not ordered.
Nonetheless the operator algebra still closes since this product
 can be rewritten as a sum of correctly ordered terms. The sum of
 terms correspond to the various commutators one picks up through
reordering ${\hat{x}}$'s and ${\hat{p}}$'s.}
. The $\star$ in the Weyl-dual function mentioned above denotes the so called
Moyal product {\cite{groe}} of two functions on phase space with respect to the standard ordering of Mehta {\cite{meht}} we are using
\footnote{This $\star$-product of phase space functions should not
 be confused with the $*$-product of elements of the fuzzy space
 ${\cal{A}}^*_n ( {\mathbb{R}}^D )$ used earlier.}
. The $\star$-product is non-commutative and associative
 (as one would expect since these properties are also true of the
operator product). The action of this product between functions can be
expressed succinctly in terms of the following exponentiated
differential operator
%
%%%     EQUATION (36)
%
\be
\star \; =\; {\mbox{exp}} \left( -i\,
{{\overleftarrow{\partial}} \over \partial p_\mu}
{{\overrightarrow{\partial}} \over \partial x^\mu} \right) \; ,
\label{eq:36}
\ee
where a left (right) pointing arrow denotes the action of
that derivative on the function to the left (right) of the
$\star$-product only
\footnote{We thank C. Zachos for pointing out an error in the expression ({\ref{eq:36}}) in an earlier version of the paper.}
. More specifically, given two functions
 ${\tilde{A}}$ and ${\tilde{B}}$ then
%
%%%     EQUATION (36a)
%
\be
{\tilde{A}} \star {\tilde{B}} \; =\; \sum_{m=0}^{\infty}
{1 \over m!} \, (-i)^m \, \left( {\partial \over
 \partial p_{\mu_1}} \cdot\cdot\cdot {\partial \over \partial
p_{\mu_{m}}} \, {\tilde{A}} \right) \left( {\partial \over
 \partial x^{\mu_1}} \cdot\cdot\cdot {\partial \over \partial
x^{\mu_{m}}} \, {\tilde{B}} \right)  \; .
\label{eq:36a}
\ee
Notice in particular that the $m=0$ term in ({\ref{eq:36a}}) is
just the commutative classical product of functions ${\tilde{A}}
{\tilde{B}}$. The $m>0$ terms are not commutative but are invariant under the combined exchange
${\tilde{A}} \leftrightarrow {\tilde{B}}$ and $x \leftrightarrow p$.
Equation ({\ref{eq:36a}}) implies that
$x^\mu \star p_\nu = x^\mu p_\nu$ and
$p_\nu \star x^\mu = x^\mu p_\nu - i \, \delta^\mu_{\; \nu}$, thus
confirming that the $\star$-product of functions preserves the
structure of the Heisenberg algebra. It is also worth noting that
partial derivatives (with respect to $x$ or $p$) act as derivations
on the algebra of classical phase space functions with
$\star$-product since they obey the Leibnitz rule when acting
on ({\ref{eq:36a}}).

The definition ({\ref{eq:36a}}) implies that
%
%%%     EQUATION (37)
%
\be
\int dx\, dp\; ({\tilde{A}} \star {\tilde{B}}) (x,p) \; =\; \int dx\,
dp\; {\tilde{A}}^{\,\prime} (x,p) {\tilde{B}}^{\,\prime} (x,p) \; =\; \int dx\, dp\; ({\tilde{B}}
\star {\tilde{A}})(x,p)  \; ,
\label{eq:37}
\ee
where the primed phase space functions denote
${\tilde{A}}^{\,\prime} := {\mbox{exp}} \left( {i\over 2} {\partial \over \partial x^\mu} {\partial \over \partial p_\mu} \right) {\tilde{A}}$ and
${\tilde{B}}^{\,\prime} := {\mbox{exp}} \left( {i\over 2} {\partial \over \partial x^\mu} {\partial \over \partial p_\mu} \right) {\tilde{B}}$
which are just multiplied with respect to the classical product in ({\ref{eq:37}}).
Thus the trace
({\ref{eq:34}}) of the operator product ${\hat{A}} {\hat{B}}$ is indeed
symmetric, as required.

The precise form of the gauge-invariant action ({\ref{eq:29}}) is
therefore given by
%
%%%     EQUATION (38)
%
\be
-{1 \over 4}  {\mbox{Tr}} \left( {\hat{F}}_{\mu\nu}
{\hat{F}}^{\mu\nu} \right) \; =\; -{1 \over 4} \int dx\,
dp\; {\tilde{F}}^{\,\prime}_{\mu\nu} (x,p) {\tilde{F}}^{\,\prime\,\mu\nu} (x,p) \; ,
\label{eq:38}
\ee
where the function ${\tilde{F}}^{\,\prime}_{\mu\nu} := {\mbox{exp}} \left( {i\over 2} {\partial \over \partial x^\mu} {\partial \over \partial p_\mu} \right) {\tilde{F}}_{\mu\nu}$
and ${\tilde{F}}_{\mu\nu}$ is the Weyl-dual of the
operator ${\hat{F}}_{\mu\nu}$, which can be obtained using ({\ref{eq:35}}).

The formal similarity with Yang-Mills theory found thus far might
lead one to expect that the field equation ({\ref{eq:28}}) follows
as the Euler-Lagrange equation for ({\ref{eq:38}}). Indeed, varying
the action ({\ref{eq:38}}) gives ${\mbox{Tr}} \left( ( \delta
{\hat{A}}^{\nu} ) [ {\hat{D}}^\mu , {\hat{F}}_{\mu\nu} ] \right)$
which would seem to suggest the equation of motion ({\ref{eq:28}}).
This is not the case however. The obstruction is due to the fact that
the trace ${\mbox{Tr}}$ does not act diagonally on the components of
general operator products. More will be said about this subtlety in
subsection 4.2.2. We just conclude by noting that the
Euler-Lagrange equation for ({\ref{eq:38}}) consists of a particular
linear combination of all the components of $[ {\hat{D}}^\mu ,
{\hat{F}}_{\mu\nu} ]$. It is therefore less restrictive than
({\ref{eq:28}}) in the sense that solutions of ({\ref{eq:28}})
are also solutions of this field equation though the converse
statement is not necessarily true. It may be possible to
 obtain ({\ref{eq:28}}) from ({\ref{eq:38}}) via additional constraints
 but we will not explore this further here.

%%%%%%%%%%%%%%%%%%%%%%%%%%%%%%%%%%%%%%%%%%%%%%%%%%%%%%%%%%%%%%%%%%%%%%%%%%%%%%%%%%%%%%%%%%%%%%%%%%%%%%%%%%%%%%%%%%%%%%%%%%%%%%%%%%%%%%%%%%%%%%%%%%%%%%%%%%

\subsubsection{Wigner basis for integrable functions}

We will now briefly describe a particular basis for a class of
classical functions which have finite integrals over phase space
(a more detailed review of this construction is given in {\cite{zach}}).
This will show us how to restrict to the class of Weyl-dual operators for
 which the trace map $ {\mbox{Tr}}$ is well-defined. Of course, this is
necessary so that the gauge-invariant action ({\ref{eq:38}}) exists.

Consider a complete orthonormal basis of eigenfunctions $\{ \psi_a \}$ for a given
Hamiltonian $H$. To each such eigenfunction $\psi_a (x)$ on
${\mathbb{R}}^D$,
 there is an associated {\emph{Wigner function}}
%
%%%     EQUATION (wigner)
%
\begin{equation}
f_a (x,p) \; =\; {1 \over (2\pi )^D} \int dy \; \psi_{a}^*
\left( x - y \right) e^{-i y^\mu p_\mu } \psi_{a}
\left( x \right) \; ,
\label{wigner}
\end{equation}
on phase space.

One can show that such Wigner functions satisfy the orthogonality
relation $f_a (x,p) \star f_b (x,p) = (1/ 2\pi )^D \, \delta_{ab} \,
 f_b (x,p)$ with respect to the $\star$-product (following from
 the fact that the eigenfunctions are orthonormal with respect to the
${\rm{L}}^2 ( {\mathbb{R}}^D )$ inner product). Consequently the set of
 Wigner functions $\{ f_a \}$ is closed with respect to $\star$
multiplication. One can also show that ({\ref{wigner}}) implies that
 each Wigner function is integrable over phase space since
$\int dx \, dp \, f_a (x,p) =1$ (this follows from the fact the
 eigenfunctions are ${\rm{L}}^2 ( {\mathbb{R}}^D )$ normalised).

Clearly linear combinations of these Wigner functions form a vector space
with a closed $\star$-product and admit partial derivatives which obey the Leibnitz rule.
Moreover, any phase space function ${\tilde{A}}$ which has an expansion
 in terms of Wigner functions, such that ${\tilde{A}} (x,p) = \sum_a
{\tilde{A}}_a \, f_a (x,p)$, is guaranteed to be integrable over phase
space provided the set of constant coefficients $\{ {\tilde{A}}_a \}$
have a finite sum $\sum_a {\tilde{A}}_a < \infty$. Thus if we restrict
to classical functions which can be expanded in this way then the
corresponding Weyl-dual operators (obtained from (\ref{eq:33})) will
have finite traces. Imposing these restrictions guarantees the
gauge-invariant action (\ref{eq:38}) is well-defined.

An explicit realisation of the Wigner basis defined above that would
 be suitable for our purposes follows from the Hamiltonian $H =
( p_\mu p^\mu + x^\mu x_\mu )/2$ (in Euclidean signature ${\mathbb{R}}^D$)
corresponding to $D$ decoupled harmonic oscillators.
The Wigner functions $f_a$ are then each proportional to
$\left[ {\mbox{exp}} \left( {-i\over 2} {\partial \over \partial x^\mu} {\partial \over \partial p_\mu} \right) \left( e^{-2H} L_a (4H) \right) \right]$,
 where $\{ L_a \, |\, a \in {\mathbb{Z}}_+ \}$ are the Laguerre polynomial
 functions.
Each of these Wigner functions has Gaussian decay at large $x$ and $p$.
This structure is appealing from the point of view of constructing
convergent integrals though it must be understood that considering only
functions of this nature on $T^* {\mathbb{R}}^D$ is quite a severe restriction.
In particular, the set of such integrable functions on the classical phase
space ${\mathbb{R}}^{2D}$ is roughly as large as the set of {\emph{all}}
functions on ${\mathbb{R}}^D$, since the basis of the former set is in
one-to-one correspondence with complete set of eigenfunctions of the
Hamiltonian for a particle moving on ${\mathbb{R}}^D$.
In this sense, the corresponding Weyl-dual space of finite trace
operators is much smaller than the complete set of functions on
$T^* {\mathbb{R}}^D$.
Consequently the number of degrees of freedom is more apt to describe
a theory living on ${\mathbb{R}}^D$, with the use of $T^* {\mathbb{R}}^D$
viewed as a tool which allows generalisation to the non-associative theory.
 We expect other ways to construct finite trace operators exist wherein,
for example, delta-normalisable functions like $e^{ikx}$ would be permitted.
In particular, it is conceivable that there exists a much larger space of
finite trace operators that would capture more of the structure of a gauge
theory on $T^* {\mathbb{R}}^D$.

%%%%%%%%%%%%%%%%%%%%%%%%%%%%%%%%%%%%%%%%%%%%%%%%%%%%%%%%%%%%%%%%%%%%%%%%%%%%%%%%%%%%%%%%%%%%%%%%%%%%%%%%%%%%%%%%%%%%%%%%%%%%%%%%%%%%%%%%%%%%%%%%%%%%%%%%%

\subsubsection{Action in position space}

The gauge-invariant action ({\ref{eq:38}}) was expressed as an integral
over phase space.
To understand the physical properties of the extended theory it would be
desirable to see how this action looks as an integral over spacetime only.
Such an expression can be obtained as follows.
Using ({\ref{eq:35}}) allows us to express ({\ref{eq:38}}) as
%
%%%     EQUATION (38a)
%
\begin{eqnarray}
&&-{1 \over 4 (2 \pi )^{4D}} \int dx dp \, dy dq \, d y^\prime d q^\prime \;
{\mbox{exp}} \left( -{i\over 2} \, ( y^\mu q_\mu + y^{\prime \mu} q^\prime_\mu ) \right) \; \nonumber \\
&&\hspace*{2.0in} \times \, {\mbox{Tr}} \left( {\mbox{exp}} ( -i\, q_\mu ( {\hat{x}}^\mu - x^\mu ) ) \,
 {\hat{F}}_{\alpha\beta} \, {\mbox{exp}} ( -i\, y^\mu ( {\hat{p}}_\mu -
p_\mu ) ) \right) \nonumber \\
&&\hspace*{2.05in} \times \, {\mbox{Tr}} \left( {\mbox{exp}} ( -i\,
q^\prime_\mu ( {\hat{x}}^\mu - x^\mu ) ) \, {\hat{F}}^{\alpha\beta} \,
{\mbox{exp}} ( -i\, y^{\prime \mu} ( {\hat{p}}_\mu - p_\mu ) ) \right)
\nonumber \\ [.1in]
&=&-{1 \over 4 (2 \pi )^{2D}} \int dy \, dq \; {\mbox{exp}} \left( -i\, y^\mu q_\mu \right) \; \nonumber \\
&&\hspace*{1.0in} \times \, {\mbox{Tr}}
\left( {\mbox{exp}} ( -i\, q_\mu {\hat{x}}^\mu ) \,
{\hat{F}}_{\alpha\beta} \, {\mbox{exp}} ( -i\, y^\mu
{\hat{p}}_\mu ) \right) {\mbox{Tr}} \left( {\mbox{exp}}
( i\, q_\mu {\hat{x}}^\mu ) \, {\hat{F}}^{\alpha\beta} \,
{\mbox{exp}} ( i\, y^\mu {\hat{p}}_\mu ) \right) \nonumber \\ [.1in]
&=&-{1 \over 4 (2 \pi )^{D}} \int dy \, dx \, d x^\prime \;
\delta ( x - x^\prime +y) \, \langle x | \, {\hat{F}}_{\alpha\beta} \,
{\mbox{exp}} ( -i\, y^\mu {\hat{p}}_\mu ) \, | x \rangle \, \langle
x^\prime | \, {\hat{F}}^{\alpha\beta} \, {\mbox{exp}} ( i\, y^\mu
{\hat{p}}_\mu ) \, | x^\prime \rangle \nonumber \\ [.1in]
&=&-{1 \over 4 (2 \pi )^{D}} \int dy \, dx \; \langle x | \,
{\hat{F}}_{\alpha\beta} \, {\mbox{exp}} ( - i\, y^\mu
{\hat{p}}_\mu ) \, | x \rangle \, \langle x+y | \, {\hat{F}}^{\alpha\beta} \, \, {\mbox{exp}} ( i\, y^\mu
{\hat{p}}_\mu ) \, | x+y \rangle  \; .
\label{eq:38a}
\end{eqnarray}
In the second line of ({\ref{eq:38a}}) we have performed the $x$ and $p$
integrals to obtain delta functions $\delta (y+ y^\prime )$ and $\delta
(q+ q^\prime )$ which have then been integrated.
The third line of ({\ref{eq:38a}}) follows by introducing a position basis
$| x \rangle$ for the traces, on which ${\hat{x}}^\mu | x \rangle =
x^\mu | x \rangle$ and ${\hat{p}}_\mu | x \rangle = -i \partial_\mu | x \rangle$.

Given the operator expansion (of the form ({\ref{eq:32}}))
for ${\hat{F}}_{\mu\nu}$ then one can formally evaluate ({\ref{eq:38a}})
in terms of the coefficient functions $F_{\mu\nu}^{\;\;\;\,
\alpha_1 ... \alpha_s}$ by writing $i {\hat{p}}_\alpha
e^{ i y^\mu {\hat{p}}_\mu } = \partial (
e^{ i y^\mu {\hat{p}}_\mu } ) / \partial y^\alpha$.
The action ({\ref{eq:38a}}) is then proportional to
%
%%%     EQUATION (38b)
%
\begin{equation}
\sum_{s,t=0}^{\infty} { (-1)^{s+t} \over s!t!} \sum_{k=0}^{t} {t \choose k} \;
K_{\alpha_1 ... \alpha_s \beta_{k+1} ... \beta_{t}}
\int dx \; F_{\mu\nu}^{\;\;\;\, \alpha_1 ... \alpha_s} (x) \,
\partial_{\beta_{1}} ... \partial_{\beta_{k}} F^{\mu\nu \, \beta_1 ... \beta_t} (x) \; ,
\label{eq:38b}
\end{equation}
where $K_{\alpha_1 ... \alpha_s \beta_{k+1} ... \beta_t}$ are \lq
volume factors'
%
%%%     EQUATION (38c)
%
\begin{equation}
\left[ {\partial \over \partial y^{\alpha_1}} ...
{\partial \over \partial y^{\alpha_s}} {\partial \over
\partial y^{\beta_{k+1}}} ... {\partial \over \partial
y^{\beta_t}} \, \delta (y) \right]_{y=0} \; ,
\label{eq:38c}
\end{equation}
that are constant totally symmetric tensors which weight
 each term in the sum ({\ref{eq:38b}}). Since the delta
function is a symmetric function then only the even rank
tensor volume factors are non-zero and are proportional to
(totally symmetrised) tensor products of the flat metric
$\eta_{\mu\nu}$. Despite the obvious divergence of each
of these weights, provided we restrict to the Wigner basis
of integrable functions described in the previous section
then the overall sum ({\ref{eq:38b}}) is guaranteed to be finite.
This statement is consistent with the fact that the volume factors
are different for each term in the sum ({\ref{eq:38b}}), so that
one cannot simply redefine the action by an overall infinite scale
to remove the individual divergent terms
(e.g. $\delta^{\prime\prime} (0) / \delta (0)$ is still infinite).

The action ({\ref{eq:38b}}) can be expressed as a
finite sum of finite terms by regulating the distributions in
({\ref{eq:38c}}). We achieve this by introducing an ultraviolet
 cutoff $N$ in the momentum integrals defining the delta function.
In particular, we define the function $\delta_N$ on Euclidean
${\mathbb{R}}^D$ such that
%
%%%     EQUATION (38d)
%
\begin{equation}
\delta_N (x) \; :=\; \left( {N \over {\sqrt{2\pi}}} \right)^D
{\mbox{exp}} \left( -N^2 x^2 /2 \right) \; .
\label{eq:38d}
\end{equation}
This corresponds to a standard representation of the Dirac delta
function in the $N \rightarrow \infty$ limit (but also satisfies
$\int dx \, \delta_N (x) =1$ for any finite $N$). At the origin
$\delta_N (0) =  (N / {\sqrt{2\pi}} )^D$. The regulated even
rank volume factors in ({\ref{eq:38c}}) can then be written
%
%%%     EQUATION (38e)
%
\begin{equation}
\partial_{\mu_1} ... \partial_{\mu_{2r}} \, \delta_N (0) \; =\;
(-1)^r \, {(2r)! \over 2^r r!} \, N^{2r} \, \eta_{( \mu_1 \mu_2}
... \eta_{\mu_{2r-1} \mu_{2r} )} \, \delta_N (0) \; ,
\label{eq:38e}
\end{equation}
whilst the odd rank factors indeed vanish identically.
This is useful because it allows one to formally factor out the
delta function at the origin in ({\ref{eq:38b}}) to obtain the
regulated action
%
%%% EQUATION (38f)
%
\begin{eqnarray}
&& \delta_N (0) \, \sum_{s,t=0}^{\infty} { 1 \over s!t!} \sum_{k=0}^{t} {t \choose k} \; (-1)^{s+t+k \over 2} {  (s+t-k)!
 \over  2^{s+t-k \over 2} \left( {s+t-k \over 2} \right)!} \, N^{s+t-k} \label{eq:38f} \\ [.1in]
&&\hspace*{2.0in}\times \, \int dx \, F_{\mu\nu}^{\;\;\;\;
 \alpha_1 ... \alpha_{s}} \, \partial_{\beta_1} ... \partial_{\beta_k} F^{\mu\nu \, \beta_{1} ... \beta_{t}}
\; \eta_{( \alpha_1 \alpha_2} ... \eta_{\alpha_{s} \beta_{k+1}} ... \eta_{\beta_{t-1} \beta_{t} )} \; , \nonumber
\end{eqnarray}
where the sums are over all terms with $s+t-k$ even.
A finite action in position space will be obtained from (\ref{eq:38f}) in the $N \rightarrow \infty$ limit where the cutoff is removed
following a simple field redefinition that will be described in section 4.3.

The expression above illustrates the point made about the
non-diagonal action of the trace on operator products at the end
of section 4.2. That is, the total symmetrisation of all the
metric indices above implies that, even in the $k=0$ sum, one does not have only diagonal
terms of the form $F_{\mu\nu \, \alpha_1 ... \alpha_{s}} \,
F^{\mu\nu \, \alpha_1 ... \alpha_{s}}$ in the Lagrangian.
Off-diagonal terms involving traces of individual field strength
component indices, like $F_{\mu\nu \, \alpha_1 ... \alpha_{s}
\alpha}^{\quad\quad\quad\;\;\; \alpha} \, F^{\mu\nu \, \alpha_1
... \alpha_{s}}$, are also present. This property is clearly
independent of having introduced the cutoff and would indeed occur
when formally evaluating ({\ref{eq:38b}}).

%%%%%%%%%%%%%%%%%%%%%%%%%%%%%%%%%%%%%%%%%%%%%%%%%%%%%%%%%%%%%%%%%%%%%%%%%%%%%%%%%%%%%%%%%%%%%%%%%%%%%%%%%%%%%%%%%%%%%%%%%%%%%%%%%%%%%%%%%%%%%%%%%%%%%%%%%%%

\subsection{The extended theory in component form}

To investigate the connection with interacting higher spin gauge
theory, it will now be enlightening to examine in more detail some of
the features of the extended theory in component form.

The components of the gauge field ${\hat{A}}_\mu$ transform
infinitesimally as
%
%%%     EQUATION (39)
%
\begin{eqnarray}
\delta A_{\mu}^{\;\; \alpha_1 ... \alpha_s} &=& -\, \partial_{\mu} \,
 \epsilon^{\alpha_1 ... \alpha_s} + {\hat{E}}^{\alpha_1 ... \alpha_s}
( {\hat{A}}_\mu ,{\hat{\epsilon}} ) \label{eq:39} \\ [.1in]
&=& -\, \partial_{\mu} \, \epsilon^{\alpha_1 ... \alpha_s} +
\sum_{k=0}^{s} \, {s \choose k} \, \sum_{r=1}^{\infty} {1 \over r!} \,
\left\{ \left( \partial_{\beta_1} ... \partial_{\beta_r} A_{\mu}^{\;\;
( \alpha_1 ... \alpha_k} \right) \epsilon^{\alpha_{k+1} ... \alpha_s )
\beta_1 ... \beta_r}  \right. \nonumber \\
&&\hspace*{2.2in} \left. - \left( \partial_{\beta_1} ... \partial_{\beta_r}
\epsilon^{( \alpha_1 ... \alpha_k} \right) A_{\mu}^{\;\; \alpha_{k+1} ...
\alpha_s ) \beta_1 ... \beta_r} \right\} \nonumber \; ,
\end{eqnarray}
under ({\ref{eq:24}}). The second line of ({\ref{eq:39}}) follows from
equation ({\ref{eq:A5}}) proven in Appendix A.

The components of the field strength ${\hat{F}}_{\mu\nu}$ ({\ref{eq:26}})
are functions which can be written in terms of the components of
${\hat{A}}_\mu$ as
%
%%%     EQUATION (40)
%
\begin{eqnarray}
F_{\mu\nu}^{\;\;\;\; \alpha_1 ... \alpha_s} &=& 2\, \partial_{[ \mu}
A_{\nu ]}^{\;\;\, \alpha_1 ... \alpha_s} - {\hat{E}}^{\alpha_1 ... \alpha_s}
( {\hat{A}}_\mu ,{\hat{A}}_\nu ) \label{eq:40} \\ [.1in]
&=& 2\, \partial_{[ \mu} A_{\nu ]}^{\;\;\, \alpha_1 ... \alpha_s} -
\sum_{k=0}^{s} \, {s \choose k} \, \sum_{r=1}^{\infty} {1 \over r!} \,
\left\{ \left( \partial_{\beta_1} ... \partial_{\beta_r} A_{\mu}^{\;\;
( \alpha_1 ... \alpha_k} \right) A_{\nu}^{\;\; \alpha_{k+1} ... \alpha_s )
\beta_1 ... \beta_r}  \right. \nonumber \\
&&\hspace*{2.2in} \left. - \left( \partial_{\beta_1} ... \partial_{\beta_r}
A_{\nu}^{\;\; ( \alpha_1 ... \alpha_k} \right) A_{\mu}^{\;\;
\alpha_{k+1} ... \alpha_s ) \beta_1 ... \beta_r} \right\} \nonumber \; ,
\end{eqnarray}
where (square) bracketed indices are
(anti)symmetrised with weight 1. By construction, the functions
$F_{\mu\nu}^{\;\;\;\, \alpha_1 ... \alpha_s}$ transform in the
($GL(D,{\mathbb{R}})$-reducible) tensor product representation
corresponding to a two-form times a totally symmetric rank-s tensor
 of the Lorentz group (written $(1,1) \otimes (s)$
\footnote{We take $( p_1 ,..., p_k )$ to denote the
$GL(D,{\mathbb{R}})$-irreducible representation corresponding to
a Young tableau with $k$ rows, each of length $p_i$ (where $i=1,...,k$).}
).

The non-linear terms in the field strength components above imply that the gauge-invariant
 action ({\ref{eq:38f}}) is not conformally-invariant. Since the momentum cutoff $N$ has length
 dimension $-1$ then the action ({\ref{eq:38f}}) is only dimensionless
 provided the field strength component $F_{\mu\nu}^{\;\;\;\, \alpha_1 ... \alpha_s}$ in ({\ref{eq:40}})
has length dimension $s$ (coordinates $x^\mu$ have dimension $1$ and metric components $\eta_{\mu\nu}$
are dimensionless). One must then introduce dimensionful coupling constants to
 ensure that the linear and
non-linear gauge field terms in ({\ref{eq:40}}) have the same dimension. If one
 considers the linear part of the
field strength ({\ref{eq:40}}) only then the action ({\ref{eq:38f}}) would be
 scale invariant provided each gauge
 field $A_{\mu}^{\;\; \alpha_1 ... \alpha_s}$ has length dimension $s+1$. A
non-linear term in ({\ref{eq:40}})
must then have length dimension $s+2$. The associated coupling constant $g$
 must therefore have dimension $-2$ for
any value of $s$. Dimensional consistency thus requires a factor of $g$ to
 multiply the non-linear terms in
both ({\ref{eq:39}}) and ({\ref{eq:40}}). Since the action ({\ref{eq:38f}})
 is quadratic in the field strength
components ({\ref{eq:40}}) then, in addition to the kinetic terms that are
 quadratic in the gauge fields, there
are both cubic and quartic interaction terms with couplings $g$ and $g^2$
respectively. This structure is just as
 in standard Yang-Mills theory.

The cutoff dependence of the $k=0$ terms in ({\ref{eq:38f}}) can be formally removed by a simple redefinition
$F_{\mu\nu}^{\;\;\;\; \alpha_1 ... \alpha_s} \rightarrow N^{-s -D/2} \, F_{\mu\nu}^{\;\;\;\; \alpha_1 ... \alpha_s}$
of each of the field strength components ({\ref{eq:40}}). The redefined action is then proportional to
%
%%% EQUATION (38g)
%
\begin{equation}
\sum_{s=0}^{\infty} \sum_{t \leq s} { (-1)^s (2s)!
 \over (s+t)!(s-t)! 2^s s!} \int dx \, F_{\mu\nu}^{\;\;\;\;
 \alpha_1 ... \alpha_{s+t}} \, F^{\mu\nu \, \alpha_{s+t+1} ... \alpha_{2s}}
\; \eta_{( \alpha_1 \alpha_2} ... \eta_{\alpha_{2s-1} \alpha_{2s} )} \; ,
\label{eq:38g}
\end{equation}
up to the addition of $k>0$ terms involving only inverse powers of $N$ that will vanish in the $N \rightarrow \infty$ limit when the cutoff is removed.

Since $N$ is dimensionful, the redefinition
modifies the dimensions of each gauge field and coupling in the action. In particular, the redefinition
of the field strength above follows from the redefinition $A_{\mu}^{\;\;\, \alpha_1 ... \alpha_s}
\rightarrow N^{-s -D/2} \, A_{\mu}^{\;\;\, \alpha_1 ... \alpha_s}$ (hence each $A_{\mu}^{\;\;\,
\alpha_1 ... \alpha_s}$ now has dimension $1-D/2$).
The effect of this redefinition in the gauge transformation ({\ref{eq:39}}) and field strength
 ({\ref{eq:40}}) is to modify the coupling
constants multiplying the non-linear terms. For every $s$, a non-linear term of fixed $r$ in
 ({\ref{eq:39}}) and ({\ref{eq:40}})
has the coupling $g$ mentioned above replaced by $g_{(r)} := N^{-r
-D/2} g$. The coupling $g_{(r)}$ has dimension $r-2+D/2$, which is
always positive for $D>2$.
Notice that, if $g$ is a fixed finite number, all the couplings $g_{(r)}$ vanish in the limit $N \rightarrow \infty$.
The single coupling constant $g_{(1)}$ can be kept finite and non-zero as $N \rightarrow \infty$,
by choosing $g$ to scale like $N^{1+D/2} g_f$ in this limit (for some finite parameter $g_f$).
The free theory can then be obtained by setting $g_{(1)} = g_f = 0$.
For small $g_f$, the interacting theory could be quantised as a perturbation of this free limit.

 Coupling constants
 with positive spin-dependent length dimensions, such as discussed above,
 have also been found for interacting higher spin
 theories in flat space in the earlier work {\cite{benben}}.
 The couplings $g_{(r)}$ are similar in structure to the 't Hooft
 parameters discussed in the context of holography for general brane solutions of string
 theory in {\cite{imsy}}. This is because the cutoff $N$ we have introduced to regulate
 the delta functions is related to the number of degrees of freedom of the extended theory.
It would be interesting to understand whether these couplings are also indicative of a
 holographic interpretation for the extended theory we have described.

%%%%%%%%%%%%%%%%%%%%%%%%%%%%%%%%%%%%%%%%%%%%%%%%%%%%%%%%%%%%%%%%%%%%%%%%%%%%%%%%%%%%%%%%%%%%%%%%%%%%%%%%%%%%%%%%%%%%%%%%%%%%%%%%%%%%%%%%%%%%%%%%%%%%%%%%%

\subsubsection{Higher spin symmetry}

By construction, the field strength components ({\ref{eq:40}}) transform in the appropriate
 covariant sense under ({\ref{eq:39}}). Indeed if we restrict attention to the linear terms in these
formulas then it is clear that $F_{\mu\nu \, \alpha_1 ... \alpha_s} = 2\,
\partial_{[ \mu} A_{\nu ] \alpha_1 ... \alpha_s}$ is invariant under
$\delta A_{\mu \, \alpha_1 ... \alpha_s} = -\, \partial_{\mu} \,
\epsilon_{\alpha_1 ... \alpha_s}$. However, as explained in {\cite{demhul}},
the most general first order infinitesimal transformation for a linear gauge field
in the representation $(1) \otimes (s)$ of the Lorentz group is
%
%%%     EQUATION (41)
%
\be
\delta A_{\mu \, \alpha_1 ... \alpha_s} = \partial_{\mu} \, \varepsilon_{\alpha_1 ... \alpha_s} + s\,
\partial_{( \alpha_1} \, \xi_{\alpha_2 ... \alpha_s ) \, \mu}  \; ,
\label{eq:41}
\ee
in terms of the rank-$s$ tensor parameters $\varepsilon_{\alpha_1 ... \alpha_s}$ (in the
totally symmetric $(s)$ representation) and $\xi_{\alpha_1 ... \alpha_{s-1} \, \mu}$ (in the
 $GL(D,{\mathbb{R}})$-reducible $(1) \otimes (s-1) $ representation). The corresponding linear
field strength that is invariant under ({\ref{eq:41}}) involves $s+1$ derivatives of
$A_{\mu \, \alpha_1 ... \alpha_s}$ and is given by
%
%%%     EQUATION (42)
%
\be
2^{s+1} \, \partial_{[ \alpha_1} ... \partial_{[ \alpha_s} \partial_{[ \mu} \,
A_{\nu ] \, \beta_s ] ... \beta_1 ]}  \; .
\label{eq:42}
\ee
The various sets of square brackets indicate antisymmetrisation of the $s+1$ pairs of indices
$[ \mu \nu ]$, $[ \alpha_1 \beta_1 ]$, ..., $[ \alpha_s \beta_s ]$ separately. It is therefore
 important to understand that the extended theory (as we have described it) does not realise
all the possible gauge symmetries even at the linear level. That is we have effectively taken
$\varepsilon = - \epsilon$ and $\xi =0$ in ({\ref{eq:41}}) and so were able to realise a
gauge-invariant field strength involving only one derivative of $A_{\mu \, \alpha_1 ... \alpha_s}$.
A consequence of this fact is that one has not enough gauge symmetry to fix all the components of
the higher spin fields that could give rise to states of negative norm in the quantum theory in Lorentzian signature. This is not an issue in Euclidean ${\mathbb{R}}^D$. The interpretation of the extended theory in Euclidean signature will be discussed in section 5.2.

%%%%%%%%%%%%%%%%%%%%%%%%%%%%%%%%%%%%%%%%%%%%%%%%%%%%%%%%%%%%%%%%%%%%%%%%%%%%%%%%%%%%%%%%%%%%%%%%%%%%%%%%%%%%%%%%%%%%%%%%%%%%%%%%%%%%%%%%%%%%%%%%%%%%%%%%%

\subsubsection{Gauge-fixing in Lorentzian signature}

Let us briefly review how a gauge-fixing of the theory in
Lorentzian signature could be achieved at the linear level, given
the full symmetry ({\ref{eq:41}}). For simplicity, let us first
focus on the $s=1$ component $A_{\mu\, \alpha}$ in
({\ref{eq:41}}). There are exactly $2D-1$ temporal components
$A_{0\, a}$, $A_{m\, 0}$ and $A_{0\, 0}$ of this field that lead
to non-unitarity (since they obstruct $A_{\mu\, \alpha}$ being
transverse to the timelike/null direction defining the
massive/massless little group of $SO(D-1,1)$). There are $2D$
gauge parameters $\varepsilon_\alpha$ and $\xi_\mu$ which contain
$2D-2$ spacelike components $\varepsilon_a$ and $\xi_m$ that can
be used to gauge away the components $A_{0\, a}$ and $A_{m\, 0}$.
One of the two timelike gauge parameter components
($\varepsilon_0$ or $\xi_0$) is itself removed
 via the residual $\delta \varepsilon_\alpha = \partial_\alpha \lambda$, $\delta \xi_\mu =
- \partial_\mu \lambda$ symmetry in ({\ref{eq:41}}).
The remaining timelike component can then be used to gauge away $A_{0\, 0}$.
As discussed in {\cite{demhul}}, this can be done in a Lorentz-covariant way
via a specific gauge-fixing procedure, resulting in the on-shell constraints $\partial^\mu A_{\mu \, \alpha} =0$,
 $\partial^\alpha A_{\mu \, \alpha} =0$ and $A_{\mu}^{\;\;\, \mu} =0$ on the gauge field.
The analysis above can be repeated for the general higher spin field $A_{\mu \,
\alpha_1 ... \alpha_s}$
and one again finds that all the temporal components of this gauge field can be removed
using the gauge
 parameters $\varepsilon_{\alpha_1 ... \alpha_s}$ and $\xi_{\alpha_1 ... \alpha_{s-1} \, \mu}$.
In terms of counting this can be easily seen from the fact that all the timelike components of
$A_{\mu \, \alpha_1 ... \alpha_s}$ can be written as $A_{0 \, \alpha_1 ... \alpha_s}$ and
$A_{\mu \, 0 \alpha_1 ... \alpha_{s-1}}$, which are in the same representations of the Lorentz
 group as the gauge parameters $\varepsilon_{\alpha_1 ... \alpha_s}$ and
$\xi_{\alpha_1 ... \alpha_{s-1} \, \mu}$
 respectively.
Of course, $A_{0 \, \alpha_1 ... \alpha_s}$ and $A_{\mu \, 0 \alpha_1 ... \alpha_{s-1}}$
have the temporal
elements $A_{0 \, 0 \alpha_1 ... \alpha_{s-1}}$ in common and so it would seem there
are ${D+s-2 \choose s-1}$
 more gauge parameters than temporal components of the gauge field.
However, the residual symmetries $\delta \varepsilon_{\alpha_1 ... \alpha_s} = s\,
\partial_{( \alpha_1} \lambda_{\alpha_2 ... \alpha_s )}$, $\delta
\xi_{\alpha_1 ... \alpha_{s-1} \, \mu}
= - \partial_\mu \lambda_{\alpha_1 ... \alpha_{s-1}}$ of the gauge parameters in
({\ref{eq:41}})
mean that precisely the ${D+s-2 \choose s-1}$ unaccounted for parameters can be
gauged away, thus
making the counting correct.
This gauge-fixing can be performed covariantly to yield the on-shell constraints $\partial^\mu
A_{\mu \, \alpha_1 ... \alpha_s} =0$, $\partial^{\alpha}
A_{\mu \, \alpha \alpha_1 ... \alpha_{s-1}} =0$,
 $A_{\mu \;\;\; \alpha_1 ... \alpha_{s-1}}^{\;\;\; \mu} =0$ and
$A_{\mu\, \;\;\, \alpha \alpha_1 ... \alpha_{s-2}}^{\;\;\; \alpha} =0$, for
each $A_{\mu \, \alpha_1 ... \alpha_s}$.

A unitary version of the extended theory we have described would require sufficient
constraints on $A_{\mu \, \alpha_1 ... \alpha_s}$ and $\epsilon_{\alpha_1 ... \alpha_s}$ to
compensate for the lack of $\xi_{\alpha_1 ... \alpha_{s-1} \, \mu}$ symmetry.
One obvious way this could be achieved would be to define the \lq partially gauge-fixed' path integral
as a sum over the $A_{\mu \, a_1 ... a_s}$ and $\epsilon_{a_1 ... a_s}$ components
only (that is where the \lq gauge' indices $\alpha$ are purely spacelike). Clearly the gauge symmetry $\epsilon_{a_1 ... a_s}$ is sufficient to
fix the non-vanishing timelike components $A_{0 \, a_1 ... a_s}$ of the higher spin fields. In the linear theory, such a gauge can be imposed in a Lorentz-covariant manner
via the constraints $\partial^{\alpha} A_{\mu \, \alpha \alpha_1 ... \alpha_{s-1}} =0$,
$A_{\mu \;\;\; \alpha_1 ... \alpha_{s-1}}^{\;\;\; \mu} =0$ and $A_{\mu\, \;\;\,
\alpha \alpha_1 ... \alpha_{s-2}}^{\;\;\; \alpha} =0$, which are invariant under gauge
 transformations with parameter satisfying $\partial^{\alpha}
\epsilon_{\alpha \alpha_1 ... \alpha_{s-1}} =0$ and
$\epsilon^{\alpha}_{\;\; \alpha \alpha_1 ... \alpha_{s-2}} =0$.
The constraint $\partial^\mu A_{\mu \, \alpha_1 ... \alpha_s} =0$,
however, only follows after fixing the remaining (harmonic part of the)
$\epsilon_{a_1 ... a_s}$ symmetry.
It is not clear though whether there exist appropriate generalisations of
the Lorentz-covariant constraints above for the non-linear theory.
As for massive gauge theories, it is possible that the symmetry broken by interactions in the extended theory implies that such constraints follow as identities from the equation of motion ({\ref{eq:28}}).

%%%%%%%%%%%%%%%%%%%%%%%%%%%%%%%%%%%%%%%%%%%%%%%%%%%%%%%%%%%%%%%%%%%%%%%%%%%%%%%%%%%%%%%%%%%%%%%%%%%%%%%%%%%%%%%%%%%%%%%%%%%%%%%%%%%%%%%%%%%%%%%%%%%%%%%%%

\subsubsection{Comments on restoring full gauge symmetry}

Since the extended theory has a Yang-Mills type structure on $T^* {\mathbb{R}}^D$
then it is not surprising that it does not realise all the higher spin symmetries. This
 is simply because of the $U(1)$  principal bundle structure over
 $T^* {\mathbb{R}}^D$ we are implicitly using. The connection on this bundle is
 just given by the covariant derivative ${\hat{D}}_\mu$ ({\ref{eq:25}})
and so the curvature ${\hat{F}}_{\mu\nu} = [ {\hat{D}}_\mu , {\hat{D}}_\nu ]$ naturally contains a linear term involving only one derivative of the higher spin fields.
As already noted above, even at the linear level one requires a higher derivative
field strength ({\ref{eq:42}}) to realise all the higher spin symmetries.
Of course, if a fully gauge-invariant formulation of the extended theory exists then
one might expect there to exist some generalised covariant derivative that would replace
partial derivatives in ({\ref{eq:41}}) and ({\ref{eq:42}}) (followed by appropriate Young
symmetrisation if necessary).
Such a generalised covariant derivative cannot simply be
${\hat{D}}_\mu$ since, by construction, this only transforms covariantly under the
$\epsilon_{\alpha_1 ... \alpha_s}$ part of the higher spin symmetry.
All that can be said is that it must reduce to ${\hat{D}}_\mu$ in the \lq partially
broken' phase of the theory we have described.

Such a partially broken structure would perhaps be similar to what happens for free higher spin theories where the Fronsdal
equations {\cite{fron}} for totally symmetric tensor gauge fields (which are second order
in derivatives) are not invariant under the most general gauge transformation for such fields.
In particular one finds that the trace part of the gauge parameter cannot be realised as
 a symmetry of the equations of motion and so is set to zero.
To construct an action realising this traceless gauge symmetry one must then impose the constraint
that the double trace of the higher spin field vanishes.
It is now known that these Fronsdal equations can be reobtained from the fully
gauge-invariant (but non-local) field equations given in
 {\cite{frasag}}, {\cite{demhul}}.
One can also realise the trace part of the gauge symmetry by reintroducing the double trace
part of the field as a compensator field which restores the full gauge symmetry in the Fronsdal
 formalism {\cite{fron}}.
A fully gauge-invariant reformulation of the extended theory may therefore require additional compensator fields.
More will be said about this point in section 5.2.

%%%%%%%%%%%%%%%%%%%%%%%%%%%%%%%%%%%%%%%%%%%%%%%%%%%%%%%%%%%%%%%%%%%%%%%%%%%%%%%%%%%%%%%%%%%%%%%%%%%%%%%%%%%%%%%%%%%%%%%%%%%%%%%%%%%%%%%%%%%%%%%%%%%%%%%%%

\subsection{Comparison with Vasiliev theory}

The non-linearities in ({\ref{eq:39}}) and ({\ref{eq:40}}) suggest that the action
and field equations found earlier describe interactions between the infinite number of
component higher spin gauge fields $A_{\mu}^{\;\;\, \alpha_1 ... \alpha_s}$.
As we have seen, even at the free level, the formulation of higher spin gauge theories
is a rather subtle problem (and is discussed, for example in {\cite{demhul}},
{\cite{demhul2}}, {\cite{dvhen}}, {\cite{bekbou}}, {\cite{fron}}, {\cite{labmor}},
{\cite{sieg}}, {\cite{tsul}}, {\cite{frasag}}).
Perhaps not surprisingly, the formulation of gauge-invariant interacting higher spin
theories is even more complicated.
Attempts to construct such models in flat space have been considered in {\cite{benben}},
{\cite{framet}}, {\cite{berbur}}.
The only known consistent framework to describe interacting higher spin gauge theories was
developed by Vasiliev {\cite{vas1}}, {\cite{vas2}} (see also {\cite{sezsun}}).
We now give a brief summary of this construction {\cite{vas3}} in order to compare the
structure with that of (the associative limit of) the extended theory we have described above.

The approach is based on an extension of the MacDowell-Mansouri formulation of anti-de
Sitter gravity {\cite{macman}}.
Recall that the frame-like formulation of gravity can be considered as a gauging of the
Poincar\'{e} group $SO(D-1,1) \ltimes {\mathbb{R}}^D$.
The gauge fields related to translations and rotations are the vielbein $e_{\mu}^{\;\;\,
\alpha}$ and the Lorentz connection $\omega_{\mu}^{\;\; \alpha\beta} = - \omega_{\mu}^{\;\;
\beta\alpha}$ respectively.
The curvature for the vielbein and Lorentz connection are respectively proportional to the
torsion and Riemann tensors of the spacetime geometry.
Setting the torsion to zero then allows one to solve for the Lorentz connection in terms of
the vielbein in the usual way.
The MacDowell-Mansouri idea was to instead gauge the $AdS_D$ isometry group $SO(D-1,2)$.
The gauge field associated with $SO(D-1,2)$ rotations is written $\omega_{\mu}^{\;\; AB} =
- \omega_{\mu}^{\;\; BA}$.
If one fixes a timelike vector in the auxiliary space ${\mathbb{R}}^{D+1}$ then this gauge
field can be decomposed into $e_{\mu}^{\;\;\, \alpha}$ and $\omega_{\mu}^{\;\; \alpha\beta}$
(with gauge indices transverse to the fixed timelike direction).
This reduced theory does not quite correspond to the gauging of the Poincar\'{e} group
described above though since the \lq translation' generators fail to commute with each
other up to an $SO(D-1,1)$ rotation with coefficient proportional to the inverse
norm-squared of the fixed timelike vector in the auxiliary space.
In fact the reduced theory precisely describes gravity on $AdS_D$ (i.e. in the presence
of a cosmological constant which is identified with the aforementioned coefficient in the
algebra).
A particularly nice feature of this construction is that it allows one to express the
Einstein-Hilbert Lagrangian (with cosmological term) in a manifestly $SO(D-1,2)$-invariant
form, in terms of the curvature of the connection $\omega_{\mu}^{\;\; AB}$ and the fixed
timelike vector in the auxiliary space only.

Vasiliev considers an extended version of this formulation of gravity on $AdS_D$ in terms
of an infinite set of fields $\{ \omega_{\mu}^{\;\; A_1 ... A_s \, B_1 ... B_s} | \,
s \geq 0 \}$.
The gauge indices for a given field here transform in the traceless $(s,s)$ irreducible
representation of the anti-de Sitter group $SO(D-1,2)$ (i.e. they correspond to
Young tableau with two rows of equal length).
Just as in the MacDowell-Mansouri formulation of gravity, one can fix a timelike vector
in the auxiliary space (whose norm is again related to the cosmological constant) and
decompose the fields above in representations of $SO(D-1,1)$.
This results in the set of generalised vielbeins $\{ e_{\mu}^{\;\;\, \alpha_1 ... \alpha_s} |
 \, s \geq 0 \}$ and Lorentz connections $\{ \omega_{\mu}^{\;\; \alpha_1 ... \alpha_s \,
\beta_1 ... \beta_t} | \, 0<t \leq s \, , \, s \geq 0 \}$.
The underlying infinite-dimensional extension of the anti-de Sitter algebra that determines
the non-linear terms in the curvature for this gauging is called $hu(1| \, sp(2) \, [D-1,2])$
 and has also been found as an extension of the conformal algebra in $D-1$ dimensions by
 Eastwood {\cite{east}} in considering all possible symmetries of the Laplace equation for
a scalar field on ${\mathbb{R}}^{D-1}$.
The details of this rather complicated algebra need not concern us here, sufficed to say
 that its maximal finite-dimensional subalgebra is $so(D-1,2)$.

The precise form of the non-linear interaction terms involving all the component fields
also becomes very complicated.
To simplify matters one can perform a linearised analysis around a fixed anti-de Sitter
background.
The gauge-invariant action for this free theory can be written in a MacDowell-Mansouri form
in terms of the linearised curvature for $\omega_{\mu}^{\;\; A_1 ... A_s \, B_1 ... B_s}$
and the fixed timelike vector only.
More precisely, it is a sum of MacDowell-Mansouri type actions with relative coefficients
fixed such that the only non-trivial variations come from the gauge transformations of the
fields $e_{\mu}^{\;\;\, \alpha_1 ... \alpha_s}$ and $\omega_{\mu}^{\;\;
\alpha_1 ... \alpha_s \, \beta}$.
The gauge parameters for these fields are written $\varepsilon^{\alpha_1 ... \alpha_s}$
and $\xi^{\alpha_1 ... \alpha_s \, \beta}$ respectively (and can be understood as
generalisations of linearised diffeomorphisms and local Lorentz transformations).
A generalisation of the \lq no torsion' constraint further implies that all
$\omega_{\mu}^{\;\; \alpha_1 ... \alpha_s \, \beta}$ can be solved in terms of (first derivatives of) the fields $e_{\mu}^{\;\;\, \alpha_1 ... \alpha_s}$.
Consequently, each remaining $\xi^{\alpha_1 ... \alpha_s \, \beta}$ symmetry can
be used to gauge away the \lq hook' part of $e_{\beta \, \alpha_1 ... \alpha_s}$
\footnote{That is, the piece of $e_{\beta \, \alpha_1 ... \alpha_s}$ which is not
totally symmetric and so transforms in the $(s,1)$ representation of the local Lorentz group.
This is simply the generalisation of the antisymmetric part of the vielbein being
removed by local Lorentz symmetry to give the graviton.}
. The resulting action is then simply an infinite sum of Fronsdal actions {\cite{fron}}
for all possible free totally symmetric fields $e_{( \mu \, \alpha_1 ... \alpha_s )}$
on $AdS_D$.
Since $e_{\mu \, \alpha}^{\quad\, \alpha \alpha_2 ... \alpha_s} =0$ then the symmetric
parts $e_{( \mu \, \alpha_1 ... \alpha_s )}$ obey the double tracelessness constraint
in {\cite{fron}}.

Despite the fact that we have considered a flat rather than anti-de Sitter background,
it is still tempting to naively identify the components of the extended gauge field
$A_{\mu}^{\;\;\, \alpha_1 ... \alpha_s}$ with the generalised vielbeins
$e_{\mu}^{\;\;\, \alpha_1 ... \alpha_s}$ in the Vasiliev theory.
The similarity between these fields does not seem to extended much beyond
their index structure though.
Even at this level there are some subtle differences.
In particular, gauge-invariance in the Vasiliev theory requires the trace
constraint $e_{\mu \, \alpha}^{\quad\, \alpha \alpha_2 ... \alpha_s} =0$, which
is not necessary in the extended theory.
As discussed in subsection 4.3.2, this constraint seems to be related to unitarity
in the extended theory.
We therefore do not exclude the possibility that a fully gauge-invariant formulation
of the extended theory could be related more closely to Vasiliev theory.
We will not investigate this possibility further here though since many of the spacetime
geometrical concepts in the Vasiliev theory (e.g. the generalised Lorentz connections)
have no obvious analogue from the gauge theory perspective.

At the level we have described it, it is not even clear whether the extended theory
contains gravity.
Indeed naively setting all the component higher spin fields (and gauge parameters)
except the vielbein-like field $A_{\mu}^{\;\;\, \alpha}$ (and parameter $\epsilon^\alpha$)
 to zero does not reproduce the frame-like formulation of gravity (that one would obtain by
 doing this for the Vasiliev theory). In particular, the only non-trivial parts of the gauge
 transformation ({\ref{eq:39}}) and field strength ({\ref{eq:40}}) then become
$\delta A_{\mu}^{\;\;\, \alpha} = - \partial_\mu \epsilon^\alpha +
 ( \partial_\beta A_\mu^{\;\;\, \alpha} ) \epsilon^\beta - ( \partial_\beta
 \epsilon^\alpha ) A_{\mu}^{\;\;\, \beta}$ and $F_{\mu\nu}^{\;\;\;\, \alpha} =2\,
 \partial_{[ \mu} A_{\nu ]}^{\;\;\, \alpha} - 2\, ( \partial_\beta A_{[ \mu}^{\;\;\,
\alpha} ) A_{\nu ]}^{\;\;\, \beta}$. This gauge transformation and field strength would
correspond to vielbein diffeomorphism and torsion tensor in the naive correspondence
 with gravity. These quantities clearly do not have the form necessary for this
correspondence to be true
\footnote{For example, unlike the vielbein, the field $A_{\mu}^{\;\;\, \alpha}$ is
 not even required to be an invertible matrix.}
(except maybe in the presence of additional constraints).

These differences should perhaps be expected since our description was considered as
 an extension of Maxwell theory which, of course, has a local $U(1)$ gauge symmetry
and a global Poincar\'{e} spacetime symmetry. The Vasiliev theory follows from a
similar extension but of gravity on $AdS_D$ with local $SO(D-1,2)$ spacetime symmetry.
 Thus, just as Vasiliev theory naturally contains anti-de Sitter gravity, we should
 expect that there exists some consistent way to embed Maxwell theory in our extended
 formalism.

%%%%%%%%%%%%%%%%%%%%%%%%%%%%%%%%%%%%%%%%%%%%%%%%%%%%%%%%%%%%%%%%%%%%%%%%%%%%%%%%%%%%%%%%%%%%%%%%%%%%%%%%%%%%%%%%%%%%%%%%%%%%%%%%%%%%%%%%%%%%%%%%%%%%%%%%%%%

\subsection{Abelian embedding}

A similarity between Vasiliev's formulation of higher spin gauge theory and the
 associative limit of the extended theory can be seen in the way one embeds a
simple Abelian gauge theory in the latter which just follows the {\emph{unfolding}}
 procedure for the former.
In particular, notice that all the non-linear terms in ({\ref{eq:39}}) and ({\ref{eq:40}})
 vanish if we take the extended gauge field ${\hat{A}}_\mu$ and parameter ${\hat{\epsilon}}$
 to have components
%
%%%     EQUATION (43)
%
\begin{eqnarray}
A_{\mu \, \alpha_1 ... \alpha_s} &=& \partial_{\alpha_1} ... \partial_{\alpha_s} \,
A_\mu \nonumber \; , \\
\epsilon_{\alpha_1 ... \alpha_s} &=& \partial_{\alpha_1} ... \partial_{\alpha_s} \,
\epsilon \label{eq:43} \; .
\end{eqnarray}
This result is proved in Appendix A where it is found that operators with components
of this form generate an Abelian subalgebra of the commutator algebra of differential operators on ${\mathbb{R}}^D$.
The constraint ({\ref{eq:43}}) relates all the higher spin components of ${\hat{A}}_\mu$
and ${\hat{\epsilon}}$ to their lowest order components $A_\mu$ and $\epsilon$.
These functions represent the Maxwell potential and parameter used in the conventional
description of Abelian gauge theory on ${\mathbb{R}}^D$.
The higher spin $s>0$ component gauge transformations ({\ref{eq:39}}) and field
strengths ({\ref{eq:40}}) are also not independent since they just correspond to $s>0$
derivatives of the $s=0$ Maxwell gauge transformation $\delta A_\mu = -\, \partial_\mu \,
\epsilon$ and field strength $F_{\mu\nu} = 2\, \partial_{[ \mu} A_{\nu ]}$.
In a similar manner, the field equations ({\ref{eq:28}}) just reduce to the Maxwell
equations $\partial^\mu F_{\mu\nu} =0$ upon imposing ({\ref{eq:43}}). The field
equations that follow from the gauge-invariant action ({\ref{eq:38g}}) upon imposing
({\ref{eq:43}}) however take the form $\left( \sum_{s=0}^{\infty} c_s \, \square^s \right)
 \partial^\mu F_{\mu\nu} =0$, where $c_s$ are constants and $\square := \partial_\mu
\partial^\mu$. As already observed for the non-linear theory, this equation is indeed
less restrictive than $\partial^\mu F_{\mu\nu} =0$. In fact, equations of this less
restrictive kind were also described for more general free higher spin fields in
{\cite{demhul}}.

The constraints ({\ref{eq:43}}) are a simple case of the more general principle of
unfolding used by Vasiliev (see e.g. {\cite{vas3}}) to show how a higher spin theory
with an infinite number of fields can describe a finite number of physical degrees of freedom.
A classic example is where one has an infinite set of independent totally symmetric
traceless tensors of increasing rank $\{ \phi_{\mu_1 ... \mu_s} | s \geq 0 \}$.
This set simply describes a massless free scalar field $\phi$ if supplemented with
the infinite set of constraints $\phi_{\alpha_1 ... \alpha_s} = \partial_{\alpha_1} ...
\partial_{\alpha_s} \, \phi$.
It is clear that such constraints relate all the higher spin fields to the scalar $\phi$.
The dynamics of this scalar field are encapsulated by the tracelessness of each
$\phi_{\mu_1 ... \mu_s}$ when supplemented with the constraints.
In particular, the tracelessness of the constraint $\phi_{\mu\nu} = \partial_\mu
\partial_\nu \, \phi$ implies that $\phi$ satisfies the massless Klein-Gordon
equation $\square\, \phi =0$.
This is similar to what we find for the embedding described above.
In particular, if $A_{\mu \, \alpha_1 ... \alpha_s}$ are traceless (for any pair
of contracted indices), then the first constraints in ({\ref{eq:43}}) are equivalent
to the correct Maxwell equations $\square \, A_\mu =0$ and $\partial^\mu A_\mu =0$
(in Lorentz gauge).
If $\epsilon_{\alpha_1 ... \alpha_s}$ are also traceless then the second constraints
in ({\ref{eq:43}}) imply $\square \, \epsilon =0$ which gives the class of harmonic
gauge transformations $\delta A_\mu = -\, \partial_\mu \, \epsilon$ under which the
aforementioned (gauge-fixed)
Maxwell equations are invariant.

%%%%%%%%%%%%%%%%%%%%%%%%%%%%%%%%%%%%%%%%%%%%%%%%%%%%%%%%%%%%%%%%%%%%%%%%%%%%%%%%%%%%%%%%%%%%%%%%%%%%%%%%%%%%%%%%%%%%%%%%%%%%%%%%%%%%%%%%%%%%%%%%%%%%%%%%%%

\subsubsection{Symplectic transformations and Abelian embedding}

The construction of the ansatz ({\ref{eq:43}}) relied on the existence of an Abelian subalgebra of
the commutator algebra of differential operators on ${\mathbb{R}}^D$.
The Abelian subalgebra used in ({\ref{eq:43}}) was the one given in Appendix A.
A generalisation which leads to other (less restrictive) ansatze can be derived as follows.

Recall that the Heisenberg algebra $[ {\hat{x}}^\mu , {\hat{p}}_\nu ] = i \delta^\mu_\nu$ in $D$-dimensions is invariant under the linear action of the symplectic group $Sp(D)$.
This group consists of  real $2D$$\times$$2D$ matrices $M$ which obey
$M J M^{t} = J$, where
%
%%%     EQUATION (44)
%
\begin{equation}
J \; =\;  \left( \matrix{0 & 1_D \cr - 1_D & 0} \right) \; ,
\label{eq:44}
\end{equation}
is the canonical symplectic form. When divided into $D$$\times$$D$ blocks,
%
%%%     EQUATION (45)
%
\begin{equation}
M \; =\; \left( \matrix{a & b \cr c & d} \right) \; ,
\label{eq:45}
\end{equation}
where $a$, $b$, $c$ and $d$ are real $D$$\times$$D$ constant matrices which obey $a b^t =
 b a^t$, $c d^t = d c^t$ and $a d^t - b c^t = 1_D$.
Block triangular symplectic matrices  of the form
({\ref{eq:45}}) with $c=0$ form a subgroup of $Sp(D)$.
The symplectic constraints for this subgroup are that $a b^t = ( a b^t )^t$ and
$d^t = a^{-1}$.
The fact that $a b^t$ is a symmetric $D$$\times$$D$ matrix for this subgroup implies
that $a^{-1} b$ is also symmetric.

Consider now the canonical position space representation of this Heisenberg algebra
(where ${\hat{x}}^\mu$ and ${\hat{p}}_\nu$ act as $x^\mu$ and $-i \partial_\nu$
respectively on the basis vectors $| x^{\mu} \rangle$ of the Hilbert space).
General operators in this representation thus correspond to functions $A(x, \partial )$.
If we write the operators $x^\mu$ and $\partial^\nu$ as a $2D$-component column vector
%
%%%     EQUATION (46)
%
\begin{equation}
X \; :=\; {x \choose \partial } \; ,
\label{eq:46}
\end{equation}
then the symplectic group defined above has the simple matrix multiplication action
$X \rightarrow MX$ on this representation.
In terms of the $D$-dimensional blocks, this translates to $x \rightarrow a\, x + b\,
\partial$ and $\partial \rightarrow c\, x + d\, \partial$.
A class of mutually commuting operators are those which are diagonal with respect to
the basis $| x^{\mu} \rangle$.
That is, functions $H(x) := A(x,0)$ of $x^\mu$ only in this representation.
The subgroup of block triangular
symplectic matrices defined above
map commuting operators to commuting operators.
In particular, such triangular matrices transform
$x \rightarrow a\, x + b\, \partial$
and $\partial \rightarrow d\, \partial$, so that the
 function $H(x)$ is mapped to the function
$H(ax+b \partial)$.
All such functions $H(ax+b \partial)$ are still mutually commuting.
This fact can be verified using the Taylor expansion
%
%%%     EQUATION (47)
%
\begin{eqnarray}
H ( ax + b \partial ) &=& H( ax) + ( \partial_\mu H( ax) ) ( a^{-1} b \, \partial )^\mu +
{1 \over 2} ( \partial_\mu \partial_\nu H(ax) ) ( a^{-1} b \, \partial )^\mu
( a^{-1} b \, \partial )^\nu + \dots \nonumber \\
&=& \sum_{s=0}^{\infty} {1 \over s!} \, ( \partial_{\mu_1} ... \partial_{\mu_s} H(ax)) \,
( a^{-1} b )^{\mu_1 \nu_1} ... ( a^{-1} b )^{\mu_s \nu_s} \, \partial_{\nu_1} ...
\partial_{\nu_s} \; ,
\label{eq:47}
\end{eqnarray}
together with the fact that $a^{-1} b$ is symmetric
(where $( a^{-1} b \, \partial )^\mu \equiv ( a^{-1} b )^{\mu}_{\;\; \nu} \partial^\nu$).
Such commuting operators clearly correspond to a
restricted class of functions on
 $T^* {\mathbb{R}}^D$ but are more general than those found in Appendix A.
The discussion in  Appendix A corresponds to operators in ({\ref{eq:47}})
with $a=b= 1_D$.

The natural generalisation of the embedding ({\ref{eq:43}}) is therefore to impose
%
%%%     EQUATION (48)
%
\begin{eqnarray}
A_{\mu \, \alpha_1 ... \alpha_s} (x) &=&
( a^{-1} b)_{\alpha_1 \beta_1} ... ( a^{-1} b)_{\alpha_s \beta_s} \
\partial^{\beta_1} ... \partial^{\beta_s} \, A_\mu (x) \nonumber \; , \\
\epsilon_{\alpha_1 ... \alpha_s} (x) &=& ( a^{-1} b)_{\alpha_1 \beta_1} ...
( a^{-1} b)_{\alpha_s \beta_s} \partial^{\beta_1} ... \partial^{\beta_s} \,
\epsilon (x) \label{eq:48} \; ,
\end{eqnarray}
on all the $s>0$ components, for any constant $D$$\times$$D$ matrices $a$ and $b$
satisfying $a^{-1} b = ( a^{-1} b )^t$ (hence $a$ must also be invertible)
\footnote{The $s=0$ components must satisfy $A_\mu (x) = A^\prime_\mu (ax)$ and
$\epsilon (x) = \epsilon^\prime (ax)$ for some auxiliary fields $A^\prime_\mu$ and
$\epsilon^\prime$.
For a given $a$, if one just defines $A^\prime_\mu (x) := A_\mu ( a^{-1} x)$ and
$\epsilon^\prime (x) := \epsilon ( a^{-1} x)$ then this constraint is simply an identity.}
.
Taking $a=b$ in ({\ref{eq:48}}) reproduces the embedding ({\ref{eq:43}}) whilst
 taking $b=0$ in ({\ref{eq:48}}) gives the embedding proposed in {\cite{ram1}}.
Since operators with components of this form commute with each other then Maxwell
theory follows in the same way as was described in the previous subsection.

%%%%%%%%%%%%%%%%%%%%%%%%%%%%%%%%%%%%%%%%%%%%%%%%%%%%%%%%%%%%%%%%%%%%%%%%%%%%%%%%%%%%%%%%%%%%%%%%%%%%%%%%%%%%%%%%%%%%%%%%%%%%%%%%%%%%%%%%%%%%%%%%%%%%%%%%%%%

\section{Comments and discussion}

This section outlines how the theory described in section 4 is related to several constructions familiar in string theory. Various generalisations of the extended theory are also discussed.
 We begin by showing how the associative limit of the extended theory is equivalent
to a gauge-invariant projection of a non-commutative gauge theory. We explain  that
consideration of the theory in Euclidean signature is important from the perspective
of fuzzy sphere solutions of Matrix theory. We suggest how a formulation in de Sitter
space could resolve the issues regarding non-unitarity in the Lorentzian theory. We then
 discuss how one might describe certain physical properties of the non-associative theory.

%%%%%%%%%%%%%%%%%%%%%%%%%%%%%%%%%%%%%%%%%%%%%%%%%%%%%%%%%%%%%%%%%%%%%%%%%%%%%%%%%%%%%%%%%%%%%%%%%%%%%%%%%%%%%%%%%%%%%%%%%%%%%%%%%%%%%%%%%%%%%%%%%%%%%%%%%%

\subsection{Relation to non-commutative gauge theory}

We have observed that the gauge theory related to the non-associative space
 ${\cal{A}}^*_n ( {\mathbb{R}}^D )$ is naturally formulated on the deformed
cotangent bundle ${\cal{A}}^*_n ( T^* {\mathbb{R}}^D )$. This extended gauge
theory was found to remain non-trivial in the associative limit in section 4.
In this limit there are $D$ covariant derivatives ${\hat{D}}_{\mu} = \partial_{\mu} + A_{\mu}
(x, \partial )$. The structure of gauge fields here is just as one finds for gauge
 theory on the $2D$-dimensional non-commutative space $T^*{\mathbb{R}}^D $ (see for example
{\cite{dounek}}, {\cite{seibwit}}), where the non-commutativity parameter is given by the
 canonical symplectic form on the cotangent bundle. One difference is that the non-commutative
gauge theory on $T^* {\mathbb{R}}^D$ has $2D$ covariant derivatives
${\hat{D}}_M = \partial_M + A_M (X)$ (where $M=1,...,2D$). The $2D$ coordinates on
$T^* {\mathbb{R}}^D$ are written $X^M = ( x^\mu, x^{\tilde{\nu}} ) = ( -i \partial^{\tilde{\mu}} , -i \partial^\nu )$
(where $\mu , \nu =1,...,D$ and ${\tilde{\mu}} = \mu + D = D+1,...,2D$) and
obey the algebra $[ X^M , X^N ] = i\, \Theta^{MN}$ (where only $\Theta^{\mu {\tilde{\nu}}}
= - \Theta^{{\tilde{\nu}} \mu} = \delta^{\mu\nu}$ are non-vanishing).
 The discussion can be formally developed for the
Lorentzian case by replacing $\delta^{\mu\nu}$ with $\eta^{\mu\nu}$, but will be more subtle since there are then two timelike directions.

A gauge field satisfying  ${\hat{D}}_{\tilde \mu } = 0$
is given by $( {\hat{A}}_\mu , {\hat{A}}_{\tilde{\nu}} ) = ( a_{\mu} , - i x_\nu )$.
This configuration has vanishing field strength
$[ {\hat{D}}_M , {\hat{D}}_N ] =0$ for any constant $a_\mu$.
 It evidently solves the field equations although for
 actions usually considered in non-commutative gauge theory
 (e.g. {\cite{harv}}), which contain the square of a shifted field strength
 $[ {\hat{D}}_M , {\hat{D}}_N ] + i\, \Theta_{MN}$, such solutions have infinite energy.
 General fluctuations around such a background in the non-commutative theory would
 not preserve the ${\hat{D}}_{\tilde{\mu}} =0$ condition.
Imposing this condition is therefore a projection of the non-commutative theory.
 This projection is consistent with the extended theory we have described in the sense
 that ${\hat{D}}_{\tilde{\mu}} =0$ is gauge-invariant under conjugation by a general
 element ${\mbox{exp}} ( \epsilon (x, \partial ) )$.
 The non-vanishing component of the field strength $[ {\hat{D}}_M , {\hat{D}}_N ]$
 for a gauge field with ${\hat{A}}_{\tilde{\nu}} = - i x_\nu$ is precisely the field strength
${\hat{F}}_{\mu\nu}$ ({\ref{eq:26}}) of the extended theory.
 The gauge-invariant action obtained in section 4.2 as an integral over position and
 momentum space is related to the standard non-commutative Yang-Mills
 action by the projection above.

%%%%%%%%%%%%%%%%%%%%%%%%%%%%%%%%%%%%%%%%%%%%%%%%%%%%%%%%%%%%%%%%%%%%%%%%%%%%%%%%%%%%%%%%%%%%%%%%%%%%%%%%%%%%%%%%%%%%%%%%%%%%%%%%%%%%%%%%%%%%%%%%%%%%%%%%%

\subsection{Matrix theory realisations}

Recall from subsection 4.3.2 that the gauge symmetry realised by the extended theory in Minkowski
 spacetime is not sufficient to remove all negative norm states from the spectrum
 (without additional constraints). This issue does not arise in Euclidean ${\mathbb{R}}^D$
 where the theory defines a statistical mechanical model of fields transforming in higher
 spin representations of $SO(D)$. A time direction can be added to this theory to obtain a
 non-relativistic system in $D+1$ dimensions with spatial $SO(D)$ invariance. One can then
consider a sphere $S^{D-1}$ embedded in the Euclidean subspace ${\mathbb{R}}^D$. This
 construction is natural in the context of fuzzy spheres which give rise {\cite{ram1}},
 {\cite{ram2}} to the class of non-associative algebras we started the paper with. These
 fuzzy sphere constructions follow as solutions to 0-brane actions (which initially
contain a time direction) as used in the BFSS Matrix theory conjecture {\cite{bfss}}.
 The fuzzy spherical worldvolume emerges from matrix degrees of
 freedom associated with the spatial directions of the spacetime only. The fuzzy
 ${\mathbb{R}}^D$ considered above is to be understood as the Euclidean embedding space
 of these fuzzy spheres.

The non-associative Lorentzian theory could also arise from a Matrix theory construction
where the matrix components are associated with a Lorentzian subspace of the physical
 spacetime. For example, one could attempt to follow the fuzzy sphere construction to
obtain a fuzzy de Sitter space from the IKKT Matrix model {\cite{ikkt}}. In fact, such
 a formulation could potentially restore full gauge-invariance in the associative limit
 of the Lorentzian theory. The reason is because one classifies positive energy states in $dS_D$ locally
\footnote{Of course, globally, de Sitter space is not causally complete and so this analysis
 is restricted to a given causal patch.}
via unitary irreducible representations of $SO(D,1)$ (rather than of $SO(D-1,1)$ in Minkowski space).
This means that one cannot always describe higher spin field theories on flat and curved spacetimes
 in terms of the same number of degrees of freedom {\cite{brivas}}.
In particular, for massless theories, gauge symmetry often requires one to introduce compensator
 fields which are coupled to the fundamental higher spin field in curved space and become
decoupled only in the flat space limit
\footnote{For example, in {\cite{brivas}} it was found that the gauge-invariant formulation
of a massless \lq hook' field (in the $(2,1)$ representation of the local Lorentz group)
 in anti-de Sitter space requires an additional \lq graviton-like' compensator field
(in the $(2)$ representation of the local Lorentz group).
The hook and graviton-like fields are only decoupled in flat space where a fully
 gauge-invariant description of each massless free field is possible.}
. Of course, for massive theories, gauge symmetry implies that the compensator fields must not decouple in flat space {\cite{zino}}, {\cite{pdm}}.
In a similar manner, it is conceivable that the gauge symmetries broken by interaction terms in the extended theory could be restored via compensator fields introduced following a de Sitter space reformulation.
It is possible that such a reformulation could have interacting \lq partially massless' phases, of the kind discovered by Deser and Waldron {\cite{deswal}} for free higher spin bosonic theories in de Sitter space, which are unitary despite only realising a reduced gauge symmetry.

%%%%%%%%%%%%%%%%%%%%%%%%%%%%%%%%%%%%%%%%%%%%%%%%%%%%%%%%%%%%%%%%%%%%%%%%%%%%%%%%%%%%%%%%%%%%%%%%%%%%%%%%%%%%%%%%%%%%%%%%%%%%%%%%%%%%%%%%%%%%%%%%%%%%%%%%%

\subsection{On the non-associative theory}

Deviation from the precise Yang-Mills type structure in the non-associative theory on
 ${\cal{A}}^*_n ( T^* {\mathbb{R}}^D )$ makes the physical properties of the associative
theory on $T^* {\mathbb{R}}^D$ found in section 4 rather difficult to generalise.
The appropriate generalisation of the associative equation of motion ({\ref{eq:28}}) is
%
%%%     EQUATION (49)
%
\be
{\hat{D}}^\mu \cdot {\hat{F}}_{\mu\nu} \; =\; 0 \; .
\label{eq:49}
\ee
This certainly gauge transforms covariantly (in the non-associative sense defined in section 3)
 and also reduces to ({\ref{eq:28}}) in the associative limit.
The precise structure of the gauge-invariant action and Abelian embedding for the non-associative
theory is less clear though.

%%%%%%%%%%%%%%%%%%%%%%%%%%%%%%%%%%%%%%%%%%%%%%%%%%%%%%%%%%%%%%%%%%%%%%%%%%%%%%%%%%%%%%%%%%%%%%%%%%%%%%%%%%%%%%%%%%%%%%%%%%%%%%%%%%%%%%%%%%%%%%%%%%%%%%%%%

\subsubsection{Non-associative trace}

Recall that the construction of a gauge-invariant action for the associative theory relied on the
 existence of a well-defined symmetric trace.
An example of such a map was given in terms of a basis of integrable Wigner functions.
We see no obvious obstruction to generalising this to the non-associative case.
The explicit construction requires a choice of function that is an appropriate non-associative
 generalisation of the harmonic oscillator Hamiltonian $( -\, \square + x^2 )/2$ used in subsection 4.2.1.
The simplest \lq Hamiltonian' function on ${\cal{A}}^*_n ( T^* {\mathbb{R}}^D )$ which reduces to this
 form in the associative limit is
%
%%%     EQUATION (50)
%
\be
{1 \over 2} \, \eta^{\mu\nu} \left( - \partial_{\mu} \partial_{\nu} + x_{\mu\nu} \right)  \; .
\label{eq:50}
\ee
One must then construct a basis of eigenfunctions of this operator on ${\cal{A}}^*_n ( {\mathbb{R}}^D )$.
The groundstate of the associative Hamiltonian is simply $\psi_0 (x) = {\mbox{exp}} \left( - x^2 /2 \right)$
(with eigenvalue $D/2$).
It is not clear whether the operator ({\ref{eq:50}}) is also bounded below but a reasonable guess for the
 corresponding non-associative state would be a \lq Gaussian' of the form
%
%%%     EQUATION (51)
%
\begin{equation}
\psi_0 (x ) \; :=\; \sum_{s=0}^{\infty} {(-1)^s \over 2^s s!} \, x^{\mu_1 \mu_2 ... \mu_{2s-1} \mu_{2s}} \,
 \eta_{\mu_1 \mu_2} ... \eta_{\mu_{2s-1} \mu_{2s}} \; .
\label{eq:51}
\end{equation}
Of course, this reduces correctly in the associative limit but is not an eigenfunction of ({\ref{eq:50}})
due to non-associative corrections.
In particular, one can show that
%
%%%     EQUATION (52)
%
\begin{equation}
- {1 \over 2} \, \partial_\mu \partial^\mu \, \psi_0 (x) \; =\; f(n) \, {D \over 2} \, \psi_0 (x) - g(n) \, {1 \over 2} \,
x_\mu^{\;\;\, \mu} \, \psi_0 (z)  \; ,
\label{eq:52}
\end{equation}
where the functions $f(n) = 1 - (n-1)/4 n^3 +...$ and $g(n) = 1 + 1/n +...$ are constants on
${\cal{A}}^*_n ( {\mathbb{R}}^D )$ which both equal unity in the associative limit (the dots
indicate higher powers in the $1/n$ expansion).
The function ({\ref{eq:51}}) is therefore an exact eigenfunction of the modified Hamiltonian
$( - \partial_{\mu} \partial^{\mu} + g(n) \, x_\mu^{\;\;\, \mu} ) /2$ with eigenvalue $f(n) \, D/2$.
It is likely that there exist $1/n$-dependent modifications of the coefficients in ({\ref{eq:51}})
that make it an exact eigenfunction of ({\ref{eq:50}}).
In this manner we expect the non-associative eigenvalue problem can be solved and the corresponding
Wigner functions constructed.
Of course, it may be that one has a much larger class of integrable functions than such Wigner functions on the fuzzy space.

%%%%%%%%%%%%%%%%%%%%%%%%%%%%%%%%%%%%%%%%%%%%%%%%%%%%%%%%%%%%%%%%%%%%%%%%%%%%%%%%%%%%%%%%%%%%%%%%%%%%%%%%%%%%%%%%%%%%%%%%%%%%%%%%%%%%%%%%%%%%%%%%%%%%%%%%%

\subsubsection{Non-associative unfolding}

Recall that one of the motivations for the extended theory was the impossibility of a naive formulation
of Abelian gauge theory on ${\cal{A}}^*_n ( {\mathbb{R}}^D )$.
A more sophisticated method might be to consider the embedding ({\ref{eq:48}}) for the non-associative
extended theory.
Unfortunately, non-associative operators of this form no longer commute and the constraints ({\ref{eq:48}})
are not well-defined under gauge transformations.
All we can say is that, fundamentally, any consistent embedding must allow all the higher spin components
$A_{\mu}^{\;\;\, \alpha_1 ... \alpha_s}$ to be solved for in terms of a single component $\Phi_\mu$ in a
gauge-invariant way.

%%%%%%%%%%%%%%%%%%%%%%%%%%%%%%%%%%%%%%%%%%%%%%%%%%%%%%%%%%%%%%%%%%%%%%%%%%%%%%%%%%%%%%%%%%%%%%%%%%%%%%%%%%%%%%%%%%%%%%%%%%%%%%%%%%%%%%%%%%%%%%%%%%%%%%%%%

\section{Summary and outlook}

We analysed gauge theory on a class of fuzzy spaces which correspond to particular non-associative deformations ${\cal{A}}^*_n ( {\mathbb{R}}^D )$ of flat spacetime ${\mathbb{R}}^D$. Gauge theory on ${\cal{A}}^*_n ( {\mathbb{R}}^D  )$ is most naturally formulated in terms of the non-associative deformation ${\cal{A}}^*_n (T^* {\mathbb{R}}^D )$ of the cotangent bundle of ${\mathbb{R}}^D$. We have extended the discussion in {\cite{ram1}} to give explicit formulas for global gauge transformations and field strengths appropriate to describe this non-associative gauge theory. The theory we considered remains non-trivial in the associative limit and can be interpreted as an interacting theory involving an infinite number of higher spin fields on ${\mathbb{R}}^D$. We have examined the physical properties of this limit of the theory in some detail.

In principle, the non-associative deformation we defined can be considered for any (pseudo-)Riemannian manifold $M$. This would proceed by deforming the realisation of such geometries as algebraic curves in flat spaces of suitably large dimension. For Euclidean signature spherical spaces, we explained that the non-associative gauge theory can be related to constructions in Matrix theory via fuzzy spheres. The associative limit of the gauge theory on Minkowski space encounters subtleties related to removing all negative norm states from the spectrum by gauge-fixing. We discussed these subtleties and proposed possible solutions. Gauge theory constructions on fuzzy de Sitter geometries from Matrix theory could help resolve such issues.

The extension of fields from functions on $M$ to functions on $T^* M$ also arises in Hull's discussion of $W$-gravity {\cite{hull}}. This suggests that the related $W$-geometries may allow non-associative deformations. The physical interpretation of the non-associativity parameter in that context is  an interesting question. A cotangent bundle construction for gauge theory on non-associative spaces has also been used in {\cite{ho}}, albeit for somewhat different reasons and for a different class of non-associative algebras. Other descriptions of non-associative gauge theories have been considered in {\cite{maji}} and {\cite{nest}} though the detailed relation to the formalism we develop here is not yet clear.

The number of higher spin fields required to describe interactions can be related to the non-associativity parameter $n$ of ${\cal{A}}^*_n ( {\mathbb{R}}^D )$. This is because, in the Matrix theory origin of these algebras, $n$ is related to the size of matrices {\cite{ram1}}. All our considerations in the present paper (starting from the validity of the derivation property of $\partial_{\mu}$) have assumed the infinite number of higher spin fields to be independent of the deformation parameter $n$. However, a more careful treatment of the Matrix theory example could allow the construction of a gauge-invariant interacting theory with finitely many higher spin fields.

%%%%%%%%%%%%%%%%%%%%%%%%%%%%%%%%%%%%%%%%%%%%%%%%%%%%%%%%%%%%%%%%%%%%%%%%%%%%%%%%%%%%%
%%%%%%%%%%%%%%%%%%%%%%%%%%%%%%%%%%%%%%%%%%%%%%%%%%%%%%%%%%%%%%%%%%%%%%

\vspace*{.4in}
{\textbf{{\large{Acknowledgments}}}}

We would like to thank Chris Hull, Prem Kumar, Bob McNees, Bill Spence, Steve Thomas, Gabriele Travaglini, Misha Vasiliev and Cosmas Zachos for useful discussions
and email correspondence.
The work of PdM is supported in part by DOE grant DE-FG02-95ER40899.
The work of SR is supported by a PPARC Advanced Fellowship held at Queen Mary, University of London.

%%%%%%%%%%%%%%%%%%%%%%%%%%%%%%%%%%%%%%%%%%%%%%%%%%%%%%%%%%%%%%%%%%%%%%%%%
%%%%%%%%%%%%%%%%%%%%%%%%%%%%%%%%%%%%%%%%%%%%%%%%%%%%%%%%%%%%%%%%%%%%%%%%%%%%%%%%%%

\section*{Appendix A : ${\hat{\sf{E}}}$}

In terms of ${\sf{E}}$ and ${\sf{F}}$ ({\ref{eq:3}}) (whose explicit form are given in Appendix B
and {\cite{ram1}}), the operator ${\hat{\sf{E}}} ( {\hat{A}},{\hat{B}} )$ ({\ref{eq:4b}}) is given by
%
%%%     EQUATION (A1)
%
\begin{eqnarray}
{\hat{\sf{E}}}({\hat{A}},{\hat{B}}) &=& \sum_{r,s=0}^{\infty} {1 \over r!s!} \; \left[ \; {\sf{E}}
( A^{\mu_1 ... \mu_r} , B^{\nu_1 ... \nu_s} )\, * \, \partial_{\mu_1} ... \partial_{\mu_r}
\partial_{\nu_1} ... \partial_{\nu_s}
\right. \nonumber \\
&&\hspace*{.7in}+ \sum_{k=1}^{s} {s \choose k} \left\{ B^{\nu_1 ... \nu_s} * ( \partial_{\nu_1} ...
\partial_{\nu_k}
A^{\mu_1 ... \mu_r} ) \right. \nonumber \\ [.1in]
&&\hspace*{1.5in}\left. - {\sf{F}}( B^{\nu_1 ... \nu_s} , ( \partial_{\nu_1} ... \partial_{\nu_k}
 A^{\mu_1 ... \mu_r} ) )
\right\} \, *\, \partial_{\nu_{k+1}} ... \partial_{\nu_s} \partial_{\mu_1} ... \partial_{\mu_r}
\label{eq:A1} \\ [.1in]
&&\hspace*{.7in}- \sum_{k=1}^{r} {r \choose k} \left\{ A^{\mu_1 ... \mu_r} * ( \partial_{\mu_1} ...
 \partial_{\mu_k}
B^{\nu_1 ... \nu_s} ) \right. \nonumber \\ [.1in]
&&\hspace*{1.5in}\left. \left. - {\sf{F}}( A^{\mu_1 ... \mu_r} , ( \partial_{\mu_1} ...
\partial_{\mu_k} B^{\nu_1 ... \nu_s} ) )
\right\} \, *\, \partial_{\mu_{k+1}} ... \partial_{\mu_r} \partial_{\nu_1} ...
 \partial_{\nu_s} \; \right] \; . \nonumber
\end{eqnarray}
This expression for ${\hat{\sf{E}}} ( {\hat{A}},{\hat{B}} )$ can be
 arranged in a derivative expansion of the form
%
%%%     EQUATION (A2)
%
\be
{\hat{\sf{E}}}({\hat{A}},{\hat{B}}) \; =\; \sum_{s=0}^{\infty} {1 \over s!} \, ( {\hat{E}}^{\,\mu_1 ... \mu_s}
 ({\hat{A}},{\hat{B}}))(x) \, * \, \partial_{\mu_1} ... \partial_{\mu_s} \; ,
\label{eq:A2}
\ee
where the coefficients ${\hat{E}}^{\,\mu_1 ... \mu_s} ({\hat{A}},{\hat{B}})$ are functions on
${\cal{A}}^*_n ( {\mathbb{R}}^D )$. The $s=0$ term in this expansion is
%
%%%     EQUATION (A3)
%
\be
{\hat{E}}({\hat{A}},{\hat{B}}) \; =\; \sum_{s=1}^{\infty} {1 \over s!} \,
\left( B^{\mu_1 ... \mu_s} *( \partial_{\mu_1} ... \partial_{\mu_s} A) - A^{\mu_1 ... \mu_s} *( \partial_{\mu_1} ...
\partial_{\mu_s} B) \right) \; .
\label{eq:A3}
\ee
Unlike in ({\ref{eq:4}}), this zeroth order term is generally non-vanishing. Notice also that it
 does not depend on ${\sf{E}}$ nor ${\sf{F}}$ and so does not vanish in the associative limit.
The $s=1$ coefficient function in ({\ref{eq:A2}}) is
%
%%%     EQUATION (A4)
%
\begin{eqnarray}
{\hat{E}}^\mu ({\hat{A}},{\hat{B}}) &=& E^\mu ( A , B ) + \sum_{s=1}^{\infty} {1 \over s!} \left\{ F^\mu
 ( A^{\nu_1 ... \nu_s} , ( \partial_{\nu_1} ... \partial_{\nu_s} B) ) - F^\mu ( B^{\nu_1 ... \nu_s} , ( \partial_{\nu_1} ...
 \partial_{\nu_s} A) ) \right\} \nonumber \\
&&+ \sum_{s=1}^{\infty} {1 \over s!} \left\{ B^{\mu \nu_1 ... \nu_s} * ( \partial_{\nu_1} ... \partial_{\nu_s} A ) -
A^{\mu \nu_1 ... \nu_s} * ( \partial_{\nu_1} ... \partial_{\nu_s} B ) \right\} \label{eq:A4} \\
&&+ \sum_{s=1}^{\infty} {1 \over s!} \left\{ B^{\nu_1 ... \nu_s} * ( \partial_{\nu_1} ... \partial_{\nu_s} A^\mu ) -
A^{\nu_1 ... \nu_s} * ( \partial_{\nu_1} ... \partial_{\nu_s} B^\mu ) \right\} \; . \nonumber
\end{eqnarray}

The operator ${\hat{\sf{E}}} ( {\hat{A}},{\hat{B}} )$ is still non-trivial in the $n \rightarrow
\infty$ limit since it reduces to the commutator $[ {\hat{B}},{\hat{A}} ]$.
The derivative expansion for this commutator can be read off from ({\ref{eq:A1}}) after taking ${\sf{E}}, {\sf{F}} \rightarrow 0$ in the associative limit.
In this limit the coefficient functions in ({\ref{eq:A2}}) are given by
%
%%%     EQUATION (A5)
%
\begin{eqnarray}
{\hat{E}}^{\,\mu_1 ... \mu_s} ({\hat{A}},{\hat{B}}) &=& \sum_{k=0}^{s} \, {s \choose k} \, \sum_{r=1}^{\infty}
 {1 \over r!} \, \left\{ ( \partial_{\nu_1} ... \partial_{\nu_r} A^{( \mu_1 ... \mu_k} ) B^{\mu_{k+1} ... \mu_s )
 \nu_1 ... \nu_r}  \right. \label{eq:A5} \\
&&\hspace*{1.2in} \left. -( \partial_{\nu_1} ... \partial_{\nu_r} B^{( \mu_1 ... \mu_k} ) A^{\mu_{k+1} ... \mu_s )
 \nu_1 ... \nu_r} \right\} \; , \nonumber
\end{eqnarray}
where bracketed indices are to by symmetrised with weight 1.

The coefficients ({\ref{eq:A5}}) generally do not vanish for any $s$.
This statement implies that the commutator algebra of differential operators on ${\mathbb{R}}^D$ is non-Abelian.
Notice though that the sum on the right hand side of ({\ref{eq:A5}}) has a symmetry under $k \rightarrow s-k$.
Therefore this commutator algebra has an Abelian subalgebra generated by operators ${\hat{H}}$ (of the form ({\ref{eq:22a}})) whose components satisfy
%
%%%     EQUATION (A6)
%
\be
H_{\mu_1 ... \mu_s} \; =\; \partial_{\mu_1} ... \partial_{\mu_s} H \; ,
\label{eq:A6}
\ee
for all $s$ (indices have been lowered using the flat metric $\eta_{\mu\nu}$). Indeed ({\ref{eq:A5}}) implies that $[ {\hat{A}},{\hat{B}} ] =0$
for any operators ${\hat{A}}$
 and ${\hat{B}}$ whose coefficient functions both take the form ({\ref{eq:A6}}).
The commutator algebra of differential operators on ${\mathbb{R}}^D$ is infinite-dimensional. Indeed an operator
of the form ({\ref{eq:22a}})
 has an infinite number of linearly independent component functions.
The constraint ({\ref{eq:A6}}) relates
 all these component fields to the zeroth order scalar function in
 the expansion of the operator.
 Consequently the Abelian subalgebra defined above is one-dimensional.

%%%%%%%%%%%%%%%%%%%%%%%%%%%%%%%%%%%%%%%%%%%%%%%%%%%%%%%%%%%%%%%%%%%%%%%%%%%%%%%%%%%%%%%%%%%%%%%%%%%%%%%%%%%%%%%%%%%%%%%%%%%%%%%%%%%%%%%%%%%%%%%%%%%%%%%%%%

\section*{Appendix B : ${\sf{E}}$ and ${\sf{F}}$}

As found in {\cite{ram1}}, the $*$-product discussed in section 2 can be written in terms of the
derivations $\partial_{\mu}$ given in  ({\ref{eq:1}}).
Writing the product in this way allows computation of its action on more
general functions. For two sets of integers $S$ and $T$, the $*$-product
 rule for basis elements of ${\cal{A}}^*_n ( {\mathbb{R}}^D )$ can be written
%
%%%     EQUATION (B1)
%
\be
x^{\mu (S)}* x^{\mu (T)} \; =\; \sum_{U \subset S,\, V \subset T } {1 \over n^{2 |U|}}
 {n! \over ( n- |U|)!} \, \partial_{\mu (U \cup V)} \, x^{ \mu ( S \cup T \setminus U \cup V )} \; ,
\label{eq:B1}
\ee
where $x^{\mu (S)} := x^{\mu_1 ... \mu_s}$ (the number of elements in $S$ is written
$|S| \equiv s$) and $\partial_{\mu (S)} := \partial_{\mu_1} ... \partial_{\mu_s}$.
One can check that the derivations $\partial_\mu$  indeed obey the Leibnitz rule when acting on ({\ref{eq:B1}}).
 The map $m_2^* : {\cal{A}}^*_n ( {\mathbb{R}}^D ) \otimes {\cal{A}}^*_n ( {\mathbb{R}}^D )
 \rightarrow {\cal{A}}^*_n ( {\mathbb{R}}^D )$ is defined such that $m_2^* ( x^{\mu(S)} \otimes
 x^{\mu(T)} ) := x^{\mu(S)} * x^{\mu(T)}$. One can also define a {\emph{concatenation}} product
map $m_2^c$ such that $m_2^c ( x^{\mu (S)} \otimes x^{\mu (T)} ) := x^{\mu (S \cup T)}$.
 We can write $m_2^*$ in terms of $m_2^c$ using the derivations such that
%
%%%     EQUATION (dervs)
%
\be
m_2^* \; =\; m_2^c \, \sum_{k=0}^{\infty} {1 \over n^{2k}} {n \choose k} \,  \partial_{\mu_1} ... \partial_{\mu_k}
\otimes \partial^{\mu_1} ... \partial^{\mu_k} \; .
\label{dervs}
\ee
This equation is of the form
%
%%%     EQUATION (intdcf)
%
\be
m_2^* \; =\; m_2^c \, f ( \partial_{\mu}  \otimes \partial^{\mu}  ) \; ,
\label{intdcf}
\ee
where the function
%
%%%     EQUATION (B2)
\be
f(x) \; :=\; \left( 1 + {x \over n^2} \right)^n \; .
\label{eq:B2}
\ee
With this expression, $f^{-1}$ is easy to write down and expand.
This allows us to recover the inverse formula expressing $m_2^c =
m_2^* \, ( f^{-1} ( \partial_{\mu} \otimes \partial^{\mu} ))$,
which agrees with (and proves) the formula in Appendix 1 of
{\cite{ram1}}. It should be noted that $\partial_{\mu}$ is also a
derivation with respect to a more general product than $m_2^*$ that can be written in terms of
$m_2^{c}$ as above but for {\emph{any}} function $f( \partial_{\mu} \otimes \partial^{\mu} )$.
If one considers a function which also depends on the degree operator acting on the
polynomials in $x^{\mu}$ then one obtains deformed derivations such as those which follow from the product $m_2$ in {\cite{ram1}}.
The remaining formulae are therefore also valid for more general products expressed as in ({\ref{intdcf}}) but for any $f$
that is a function of $\partial_{\mu} \otimes \partial^{\mu}$ only.

Using the fact that the concatenation product is associative now enables us to give expressions for the
 associativity operators ${\sf{E}}$ and ${\sf{F}}$ ({\ref{eq:3}}) in terms of the function $f$. Let us
begin by defining a map $\Delta$ from the space of multiple derivatives of ${\cal{A}}^*_n ( {\mathbb{R}}^D )$
to the tensor product of two copies of this space. That is
%
%%%     EQUATION (B3)
%
\be
\Delta ( \partial_{\mu (S)}  ) \; :=\; \sum_{U \cup V  = S} \partial_{\mu (U)} \otimes \partial_{\mu ( V )  } \; .
\label{eq:B4}
\ee
Consequently, we can write $\partial_{\mu (S)} m_2^* = m_2^* \, \Delta ( \partial_{\mu (S)} )$ when acting on
${\cal{A}}^*_n ( {\mathbb{R}}^D ) \otimes {\cal{A}}^*_n ( {\mathbb{R}}^D )$. The same equation
also holds when we replace $m_2^*$ with $m_2^c$, since $\partial_{\mu}$ also obeys the Leibnitz rule with respect to $m_2^c$.

Now consider the product $A*(B*C)$ for any three functions on ${\cal{A}}^*_n ( {\mathbb{R}}^D )$,
%
%%%     EQUATION (abc1)
%
\begin{eqnarray}
A*(B*C) &=& m_2^*(1 \otimes m_2^*)( A \otimes B \otimes C ) \nonumber \\
&=& m_2^c \, f ( \partial_\mu \otimes \partial^\mu )( 1 \otimes m_2^c) f ( 1 \otimes \partial_\nu \otimes \partial^\nu )
(A \otimes B \otimes C) \label{abc1} \\
&=& m_2^c \, ( 1 \otimes m_2^c ) ( ( 1 \otimes \Delta )f ( \partial_\mu \otimes \partial^\mu ))
f ( 1 \otimes \partial_\nu \otimes \partial^\nu )(A \otimes B \otimes C) \nonumber \; ,
\end{eqnarray}
which can be rearranged using associativity of $m_2^c$ to give
%
%%% EQUATION (B4)
%
\begin{eqnarray}
&=& m_2^* \, f^{-1} ( \partial_\mu \otimes \partial^\mu )( m_2^* \otimes 1) f^{-1} ( \partial_\nu \otimes \partial^\nu )
(( 1 \otimes \Delta )f( \partial_\rho \otimes \partial^\rho ))f(1 \otimes \partial_\sigma \otimes \partial^\sigma )
(A \otimes B \otimes C) \nonumber \\
&=& m_2^* \, ( m_2^* \otimes 1 )  ( ( \Delta \otimes 1  ) f^{-1}( \partial_\mu \otimes \partial^\mu ))
f^{-1} ( \partial_\nu \otimes \partial^\nu )(( 1 \otimes \Delta )f( \partial_\rho \otimes \partial^\rho )) \nonumber \\
&&\hspace*{3.8in} \times f( 1 \otimes \partial_\sigma \otimes \partial^\sigma )(A \otimes B \otimes C) \; . \nonumber \\
\label{eq:B4}
\end{eqnarray}
This manipulation has expressed the product $A*(B*C)$ in terms of a sum of products
involving derivatives of the functions $A$, $B$ and $C$ (where the $*$-multiplication
of the first two entries is done first and so is similar in structure to $(A*B)*C$).
Thus the ${\sf{F}}$ operator in ({\ref{eq:3}}) can be read off from
%
%%%     EQUATION (B5)
%
\begin{eqnarray}
{\sf{F}} (A,B) C &=&  m_2^* \, ( m_2^* \otimes 1 ) [ 1 - ( ( \Delta \otimes 1  )
f^{-1}( \partial_\mu \otimes \partial^\mu )) f^{-1} ( \partial_\nu \otimes \partial^\nu ) \nonumber \\
&&\hspace*{1.5in} \times (( 1 \otimes \Delta )f( \partial_\rho \otimes \partial^\rho ))
f( 1 \otimes \partial_\sigma \otimes \partial^\sigma ) ] (A \otimes B \otimes C) \; , \nonumber \\
\label{eqn:B5}
\end{eqnarray}
in terms of derivatives acting on $C$ (as in ({\ref{eq:4}})). The analogous derivative
 expansion of the ${\sf{E}}$ operator immediately follows, since ${\sf{E}}(A,B) = {\sf{F}}(A,B) -
 {\sf{F}}(B,A)$.

%%%%%%%%%%%%%%%%%%%%%%%%%%%%%%%%%%%%%%%%%%%%%%%%%%%%%%%%%%%%%%%%%%%%%%%%%%%%%%%%%%%%%%%%%%%%%%%%%%%%%%%%%%%%%%%%%%%%%%%%%%%%%%%%%%%%%%%%%%%%%%%%%%%%%%%%%%%

\end{document}